\documentclass{MYws-mpla}
\usepackage[superscript]{cite}

\def\bra#1{\langle #1 |\, }
\def\ket#1{\, | #1 \rangle}
\newcommand{\ba}{\begin{array}}
\newcommand{\ea}{\end{array}}
\newcommand{\no}{\nonumber}
\newcommand{\be}{\begin{equation}}
\newcommand{\ee}{\end{equation}}
\newcommand{\beqn}{\begin{eqnarray}}
\newcommand{\eeqn}{\end{eqnarray}}
\newcommand{\eqn}[1]{(\ref{#1})}
\newcommand{\cO}{{\cal O}}
\newcommand{\bel}[1]{\be\label{#1}}

\newcommand{\smtau}{\frac{s}{m_{\tau}^{2}}}
\newcommand{\ImPi}{\operatorname{Im}\Pi}

\begin{document}

\markboth{A. Pich, A. Rodr\'\i guez-S\'anchez}
{$\alpha_s(m_\tau^2)$ from $\tau$ decays}


\title{UPDATED DETERMINATION OF  $\alpha_s(m_\tau^2)$
FROM TAU DECAYS}

\author{\footnotesize ANTONIO PICH}

\address{Departament de F\'\i sica Te\`orica, IFIC, Universitat de Val\`encia -- CSIC,\\
Apt. Correus 22085, E-46071 Val\`encia, Spain\\
Antonio.Pich@ific.uv.es}

\author{ANTONIO RODR\'IGUEZ-S\'ANCHEZ}

\address{Departament de F\'\i sica Te\`orica, IFIC, Universitat de Val\`encia -- CSIC,\\
Apt. Correus 22085, E-46071 Val\`encia, Spain\\
Antonio.Rodriguez@ific.uv.es}

\maketitle


\begin{abstract}
Using the most recent release of the ALEPH $\tau$ decay data, we present a very detailed phenomenological update of the $\alpha_s(m_\tau^2)$ determination. We have exploited the sensitivity to the strong coupling in many different ways, exploring several complementary methodologies. All determinations turn out to be in excellent agreement, allowing us to extract a very reliable value of the strong coupling. We find $\alpha_{s}^{(n_f=3)}(m_\tau^2)  = 0.328 \pm 0.012$ which implies
$\alpha_{s}^{(n_f=5)}(M_Z^{2}) = 0.1197\pm 0.0014$.
We critically revise previous work, and point out the problems flawing some recent analyses which claim slightly smaller values.

\end{abstract}

\ccode{PACS Nos.: 12.38.-t, 12.38.Qk, 13.35.Dx, 14.60.Fg}

\section{Introduction}

This workshop contribution summarizes the updated determination of the strong coupling from $\tau$ decays,
performed recently in Ref.~\refcite{Pich:2016bdg} with a very comprehensive analysis of the most recent experimental data. The $\tau$ decay width is very sensitive to $\alpha_s$ and provides a rigorous determination of the QCD coupling at the $\tau$ mass scale.\cite{Pich:2013lsa}
Owing to its inclusive character, the total hadronic decay width of the $\tau$ lepton can be analyzed with well-understood short-distance QCD tools, such as the operator product expansion (OPE).\cite{Braaten:1991qm} It turns out to be completely dominated by the perturbative contribution,
which allows us to determine $\alpha_s(m_\tau^2)$ with very good accuracy.\cite{Pich:2013lsa,Pich:2015ivv} When evolved to higher scales, it provides one of the most precise experimental determinations of $\alpha_s(M_Z^2)$,\cite{Pich:2013sqa,d'Enterria:2015toz,Agashe:2014kda,Deur:2016tte} because the long running of the strong coupling between $m_\tau$ and $M_Z$ shrinks the error by a factor roughly proportional to $\alpha^2_s(M_Z^2)/\alpha^2_s(m_\tau^2)\sim 0.1$.

The main uncertainty in the $\alpha_s$ determination from $\tau$ decays has a perturbative origin, related to the sizable value of the strong coupling at $m_\tau$ that makes it sensitive to unknown higher-order corrections. Non-perturbative contributions are small, below 1\%, and
can be identified analyzing the invariant-mass distribution of the final hadrons in $\tau$ decays.
This kinematical distribution constitutes a precious source of information to investigate non-perturbative effects and measure the parameters characterizing the QCD vacuum.\cite{LeDiberder:1992zhd}

Using the recently updated ALEPH $\tau$ spectral functions,\cite{Davier:2013sfa} we have made an exhaustive analysis of potential non-perturbative contributions to the inclusive $\tau$ decay width, to better assess their possible impact on the determination of $\alpha_s(m_\tau^2)$. We have investigated these effects with different strategies and have performed many complementary tests. In all cases, the fitted value of $\alpha_s(m_\tau^2)$ exhibits an impressive stability, showing very little sensitivity to non-perturbative corrections. In the following, we present the main results of this analysis and derive a very precise value of the strong coupling. Additional details can be found in Ref.~\refcite{Pich:2016bdg}.

\section{Inclusive Hadronic Width of the $\tau$ Lepton}

It is convenient to normalize the hadronic decay width to the leptonic one,\cite{Narison:1988ni,Braaten:1988hc,Braaten:1991qm} 
\be\label{rtau}
R_{\tau}\; =\;\frac{\Gamma[\tau^{-}\rightarrow \nu_{\tau} + \mathrm{hadrons}]}{\Gamma[\tau^{-}\rightarrow \nu_{\tau} e^{-}\overline{\nu}_{e}]}
\; =\; R_{\tau,V} + R_{\tau,A} +R_{\tau,S} \, ,
\ee
and express the ratio through the spectral identity
\be\label{eq:RtauSpectral}
R_{\tau}\; =\; 12 \pi\, S_{\mathrm{EW}}\, \int^{m_ {\tau}^{2}}_{0}\frac{ds}{m_{\tau}^{2}}\left(1-\smtau\right)^2
\left[\left(1+2\smtau\right)\ImPi^{(1)}(s)+\ImPi^{(0)}(s)\right] ,
\ee
with
\bel{eq:Pidef}
\Pi^{(J)}(s)\;\equiv\; \sum_{q=d,s}|V_{uq}|^{2}\left(\Pi^{(J)}_{uq, V}(s)+ \Pi^{(J)}_{uq, A}(s)  \right) \, ,
\ee
where $\Pi^{(J)}_{ij, \mathcal{J}}(s)$
are the two-point correlation functions for the vector
$V_{ij}^{\mu}=\overline{q}_{j}\gamma^{\mu} q_{i}$ and axial-vector
$A_{ij}^{\mu}=\overline{q}_{j} \gamma^{\mu}\gamma_5 q_{i}$ colour-singlet quark currents:
\be
i \int d^{4}x\; e^{iqx} \;
\bra{0}T[\mathcal{J}_{ij}^{\mu}(x)\mathcal{J}_{ij}^{\nu \dagger}(0)]\ket{0}
\; =\;
(-g^{\mu\nu}q^{2}+q^{\mu}q^{\nu})\; \Pi^{(1)}_{ij,\mathcal{J}}(q^{2})
+ q^{\mu}q^{\nu}\;\Pi^{(0)}_{ij,\mathcal{J}}(q^{2}) \, .
\ee
The $|V_{ud}|^{2}$ terms in \eqn{eq:Pidef} correspond to $R_{\tau,V}$ and $R_{\tau,A}$, while $R_{\tau,S}$ contains the Cabibbo-suppressed contributions.
The factor $S_{\mathrm{EW}} = 1.0201 \pm 0.0003$ in Eq.~\eqn{eq:RtauSpectral} incorporates the renormalization-group-improved electroweak correction.\cite{Marciano:1988vm,Braaten:1990ef,Erler:2002mv}

\begin{figure}[tb]
\centerline{
\includegraphics[width=0.36\textwidth]{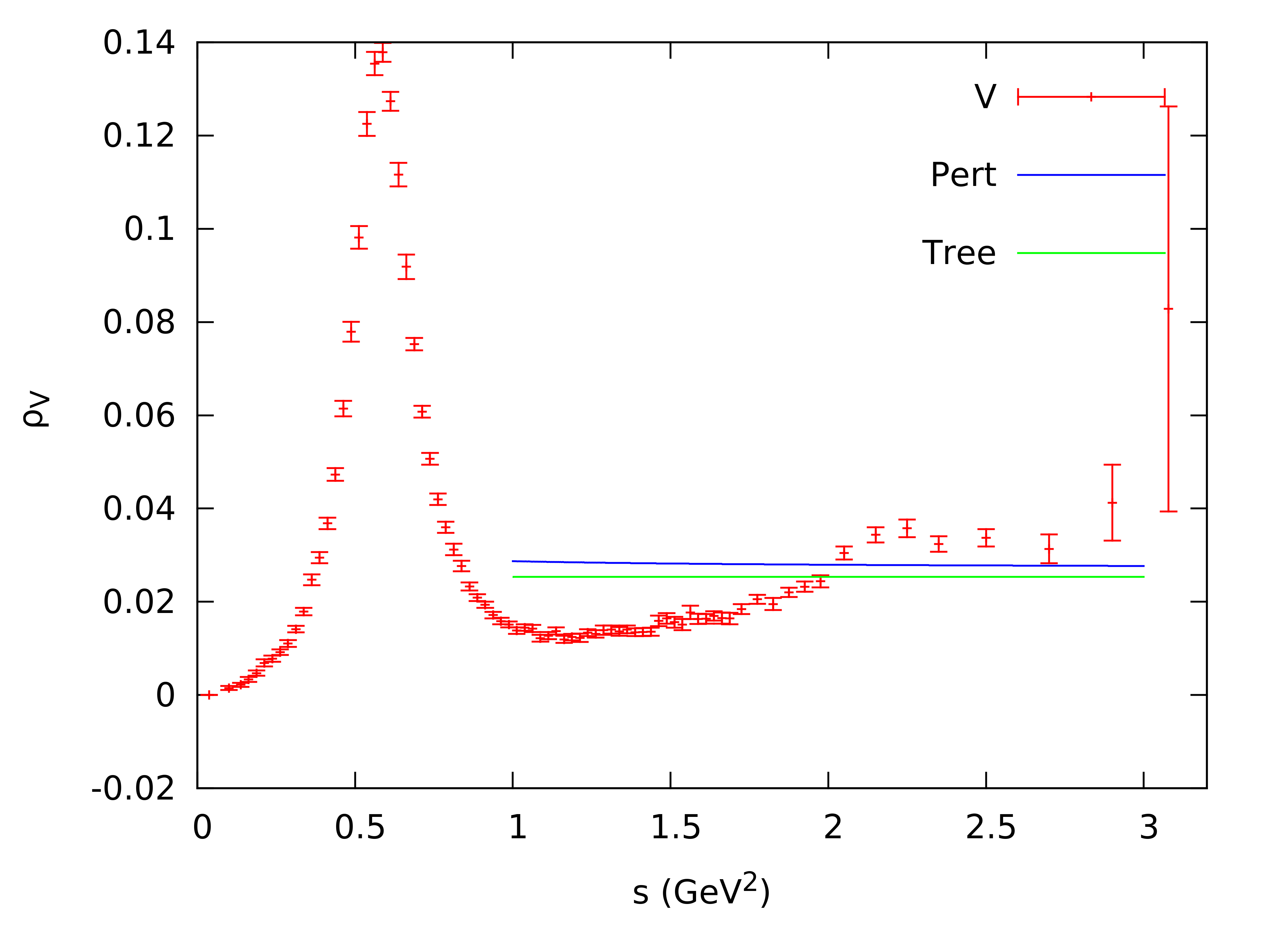}\hskip -.15cm
\includegraphics[width=0.36\textwidth]{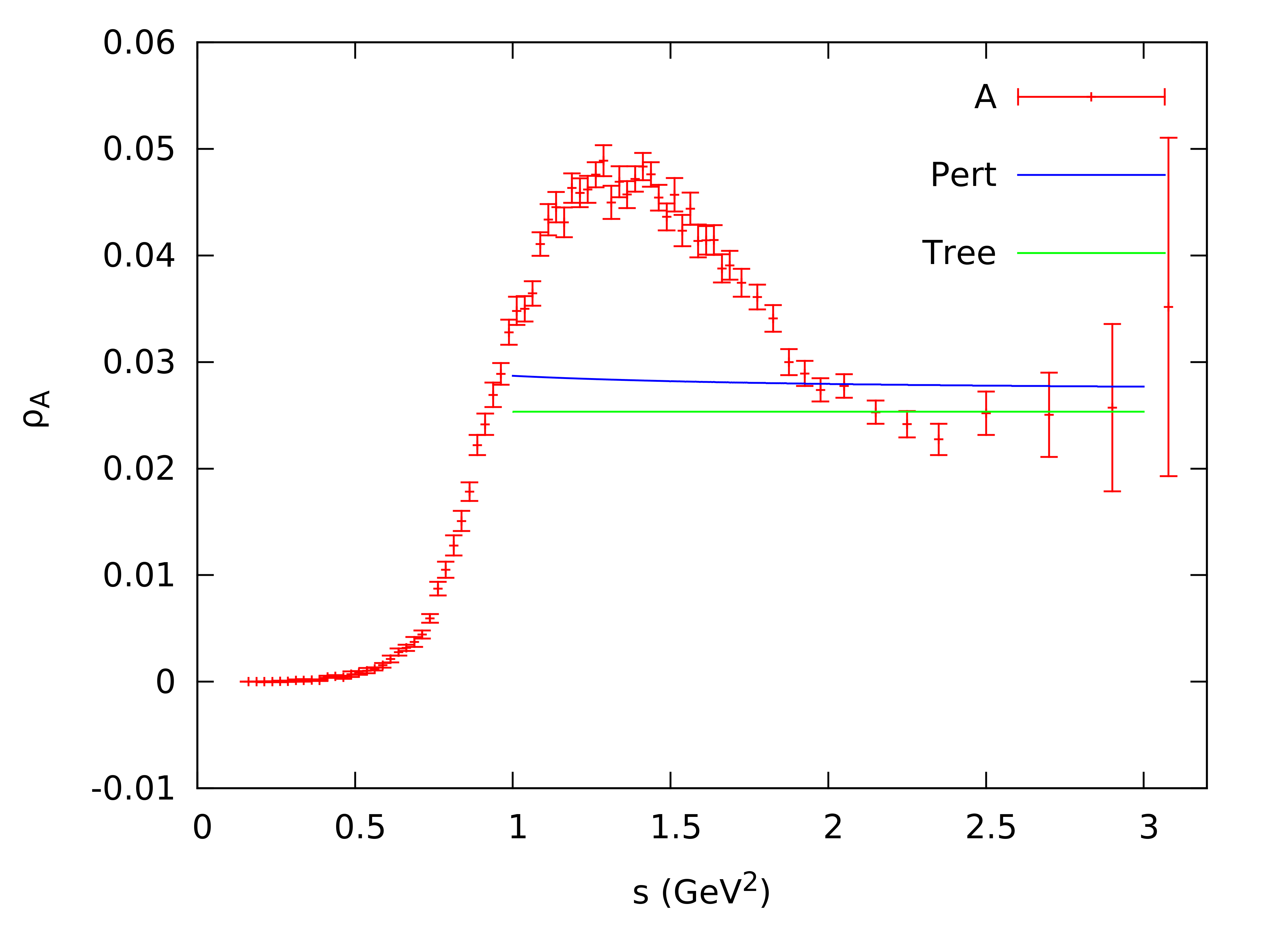}\hskip -.15cm
\includegraphics[width=0.36\textwidth]{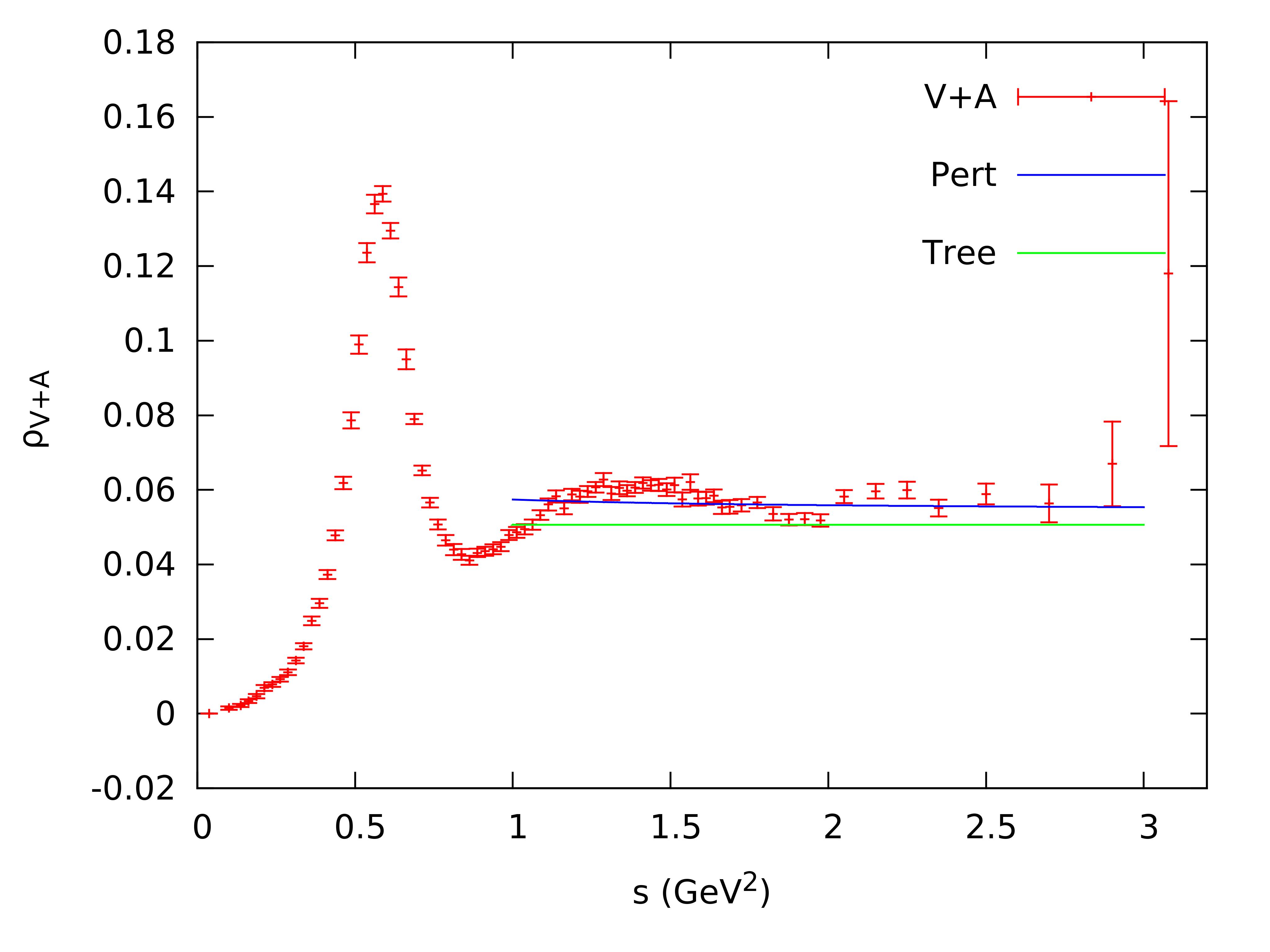}
}
\vspace*{8pt}
\caption{\protect\label{fig:ALEPHsf}
ALEPH spectral functions for the $V$, $A$ and $V+A$ channels.\protect\cite{Davier:2013sfa}}
\end{figure}

We will only discuss here the total Cabibbo-allowed width. The recently updated ALEPH spectral functions
$\rho_{ud,\mathcal{J}}(s)\equiv\frac{1}{\pi}\ImPi^{(0+1)}_{ud,\mathcal{J}}(s)$,
are shown in Fig.~\ref{fig:ALEPHsf}.\cite{Davier:2013sfa}
%
%
Using the analyticity properties of the correlators, this experimental information can be related with theoretical QCD predictions through moments of the type\cite{Braaten:1991qm,LeDiberder:1992zhd}
\be\label{aomega}
A^{\omega}_{V/A}(s_{0})\;\equiv\; \int^{s_{0}}_{s_{\mathrm{th}}} \frac{ds}{s_{0}}\;\omega(s)\, \ImPi_{V/A}(s)\; =\; \frac{i}{2}\;\oint_{|s|=s_{0}}
\frac{ds}{s_{0}}\;\omega(s)\, \Pi_{V/A}(s)\, ,
\ee
where
$\Pi_{V/A}(s)\equiv \Pi_{ud,V/A}^{(1+0)}(s)$,
$\omega(s)$ is any weight function analytic in $|s|\le s_0$, $s_{\mathrm{th}}$ is the hadronic mass-squared threshold, and the complex integral in the right-hand side (rhs) runs counter-clockwise around the circle $|s|=s_{0}$.
The OPE,
\be\label{eq:ope}
\Pi^{\mathrm{OPE}}_{V/A}(s)\; =\; \sum_{D}\frac{1}{(-s)^{D/2}}\sum_{\mathrm{dim} \, \mathcal{O}=D} C_{D, V/A}(-s,\mu)\;\langle\mathcal{O}(\mu)\rangle
\;\equiv\; \sum_{D}\;\dfrac{\mathcal{O}_{D,\, V/A}}{(-s)^{D/2}}\, ,
\ee
can be used to predict the rhs integral as an expansion in inverse powers of $s_0$, for large-enough values of $s_0$.
Differences between the physical values of the $A^{\omega}_{V/A}(s_{0})$ moments and their OPE approximations are known as quark-hadron duality violations. They are minimized by taking ``pinched'' weight functions which vanish at $s=s_0$, suppressing in this way the contributions from the region near the real axis where the OPE is not valid.\cite{Braaten:1991qm,LeDiberder:1992zhd}

\section{Perturbative Contribution}

For $s_0\sim\cO(m_\tau^2)$, the moments $A_{V/A}^{\omega}(s_{0})$ are dominated by the $D=0$ term in the OPE which contains the pure perturbative contribution for massless quarks. Owing to chiral symmetry,
the vector and the axial-vector perturbative correlators are identical. They are conveniently expressed in terms of the Adler function
\beqn\label{adler}
D(s)\;\equiv\; -s\,\frac{d\,\Pi^{P}(s)}{ds}\; =\;\frac{1}{4\pi^{2}}\;\sum_{n=0}\tilde K_{n}(\xi)\;
\left(\frac{\alpha_s(-\xi^2 s)}{\pi}\right)^n \,.
\eeqn
The coefficients $K_n\equiv \tilde K_{n}(\xi=1)$ are  known up to $n\le 4$. For $N_f=3$ flavours, one has:\cite{Baikov:2008jh}
$K_0 = K_1 = 1$, $K_2 = 1.63982$, $K_3^{\overline{\mathrm{MS}}} = 6.37101$ and
$K_4^{\overline{\mathrm{MS}}} = 49.07570$.
The homogeneous renormalization-group equation satisfied by $D(s)$ determines the corresponding scale-dependent parameters $\tilde K_{n}(\xi)$.\cite{LeDiberder:1992jjr,Pich:1999hc}

Integrating by parts the rhs of Eq.~\eqn{aomega}, the perturbative contribution to $A_{V/A}^{\omega}(s_{0})$ takes the form
\be
A^{\omega, P}(s_{0})\; =\; -\frac{1}{8\pi^{2}s_{0}}\;\sum_{n=0}\; \tilde K_{n}(\xi)\;\int^{\pi}_{-\pi} d\varphi\;
\left[ W(-s_{0}\, e^{i\varphi})-W(s_{0})\right]\; \left(\frac{\alpha_s(\xi^2 s_{0} \, e^{i\varphi})}{\pi}\right)^n , \label{pertu}
\ee
with $W(s)\equiv \int^{s}_{0}ds'\,\omega(s')$. The contour integrals multiplying the coefficients
$\tilde K_{n}(\xi)$ only depend on $\alpha_s(\xi^2 s_{0})$. They can be computed with high accuracy solving the $\beta$-function equation, up to unknown $\beta_{n >4}$ contributions. One gets in this way a {\it contour-improved perturbation theory} (CIPT) series,\cite{LeDiberder:1992jjr,Pivovarov:1991rh} which sums big running corrections arising at large values of $\varphi$, is stable under changes of the renormalization scale $\xi$ and has a very good perturbative convergence.
If one truncates instead the integrals to a fixed order in $\alpha_s(\xi^2 s_{0})$ ({\it fixed-order perturbation theory}, FOPT), the resulting series has a slow convergence and a much larger dependence on $\xi$.

\section{Sensitivity of $R_\tau$ to $\alpha_s(m_\tau^2)$}

The determination of the strong coupling from $\tau$ decays takes advantage of several properties that make $R_\tau$ particularly suitable for a precise theoretical analysis:\cite{Braaten:1991qm}
\begin{itemize}
\item[i)] The tau mass is large enough to safely use the OPE at $s_0 = m_\tau^2$.
\item[ii)] $\alpha_s(m_\tau^2)\sim 0.33$ is sizeable, making $R_\tau$ more sensitive to the strong coupling than higher-energy observables.
\item[iii)] The perturbative correction to $R_\tau$ is known to $\cO(\alpha_s^4)$ and gives a total contribution of 20\%, a quite large effect.
\item[iv)] The phase-space factor in \eqn{eq:RtauSpectral} contains a double zero at $s=m_\tau^2$ which heavily suppresses the contribution to the contour integral from the region near the real axis, where the OPE is not valid.
\item[v)] For massless quarks, $s\, \Pi^{(0)}(s)=0$. Therefore, only the correlator $\Pi^{(0+1)}(s)$ contributes to Eq.~\eqn{eq:RtauSpectral}, weighted with the function $\omega(x) = (1-x)^2 (1+2x) = 1-3x^2+2x^3$. According to Cauchy's theorem, the inclusive hadronic width is only sensitive to OPE corrections with $D=6$ and 8, which are strongly suppressed by the corresponding powers of the $\tau$ mass.
    The usually leading $D=4$ power corrections can only contribute with an additional logarithmic suppression factor of $\cO(\alpha_s^2)$, which makes their effects negligible.
\item[vi)] In addition to the $1/m_\tau^6$ suppression of non-perturbative corrections to $R_{\tau,V/A}$, there is a cancellation between the vector and axial-vector $D=6$ contributions to $R_{\tau,V+A}$, which have opposite signs.
\item[vii)] As shown in Fig.~\ref{fig:ALEPHsf}, the inclusive $V+A$ spectral distribution is very flat. The prominent $\rho(2\pi)$ and $a_1(3\pi)$ resonance structures get very soon diluted by the opening of high-multiplicity hadronic thresholds. The data approaches very fast the perturbative QCD predictions, which seem to work even at surprisingly low values of $s\sim 1.2\;\mathrm{GeV}^2$.
\end{itemize}

The large value of the strong coupling at the $\tau$ mass scale implies also that $R_\tau$ is quite sensitive to the unknown higher-order perturbative corrections, making them the largest source of uncertainty in the $\alpha_s(m_\tau^2)$ determination. For a given value of $\alpha_s(m_\tau^2)$, FOPT predicts a slightly larger perturbative correction than CIPT; therefore it leads to a smaller fitted value of $\alpha_s(m_\tau^2)$.

The numerical size of the small non-perturbative effects can be extracted from the measured invariant-mass distribution of the final hadrons, using weighted moments more sensitive to power corrections.\cite{LeDiberder:1992zhd}
The non-perturbative contribution to $R_\tau$ has been found experimentally to be safely below 1\% with the ALEPH,\cite{Davier:2013sfa,Davier:2008sk,Schael:2005am} OPAL\cite{Ackerstaff:1998yj} and CLEO\cite{Coan:1995nk} data,
in agreement with theoretical expectations.\cite{Braaten:1991qm}
Using the updated ALEPH spectral functions in Fig.~\ref{fig:ALEPHsf},
we have performed a more detailed analysis, confirming the strong suppression of non-perturbative contributions in the $V+A$ channel.\cite{Pich:2016bdg}

The structure of power corrections can be easily understood. Neglecting the higher-order logarithmic dependence of the Wilson coefficients $C_{D, V/A}$ in Eq.~\eqn{eq:ope}, $\mathcal{O}_{D,\, V/A}$ represents an effective $s$-independent vacuum condensate of dimension $D$, which can only contribute to the moment $A_{V/A}^{\omega}(s_{0})$ if the weight function $\omega(s)$ contains the power $s^{D/2-1}$. The corresponding contribution is suppressed by a factor $1/s_0^{D/2}$.
The lowest-dimensional vacuum condensates,\cite{Braaten:1991qm,Pich:2016bdg}
\bel{eq:O4}
\cO_{4,V/A}\; \approx\; \frac{1}{12\pi}\,\langle \alpha_s GG\rangle \, +\,
(m_{u}+m_{d})\,\langle\,\bar{q}q\rangle
\; \approx\; \left[(1.7\pm 0.8)\, -\, 0.16\right]\,\cdot 10^{-4}\,\times\, m_\tau^4
\, ,
\ee
are too small to provide any sizeable effect at $s_0\sim m_\tau^2$, within the much larger background noise from perturbative uncertainties and experimental errors.

\section{Updated Determination of $\alpha_s(m_\tau^2)$}

We have made a very comprehensive reanalysis of the $\alpha_s(m_\tau^2)$ determination from $\tau$ decay data, with all kinds of consistency checks to assess the potential size of non-perturbative effects.\cite{Pich:2016bdg} All strategies adopted in previous works have been investigated, studying the stability of the results and trying to uncover any potential hidden weaknesses, and several complementary approaches have been put forward. Several determinations, using different methodologies, have been performed, finding a very consistent set of results. Table~\ref{tab:summary} summarizes the
most reliable determinations, extracted from the $V+A$ channel.

All analyses have been done both in CIPT and FOPT. Within a given approach the perturbative errors
have been estimated varying the renormalization scale in the interval $\xi^2 \in (0.5\, ,\, 2)$, and taking $K_{5}=275 \pm 400$ as an educated guess of the maximal range of variation of the unknown fifth-order contribution.\cite{Pich:2011bb} These two sources of theoretical uncertainty have been combined in quadrature, together with the experimental errors. The different values quoted in the table include, as an additional uncertainty, the variations of the results under various modifications of the fit procedures. The systematic difference between the values obtained with the CIPT and FOPT prescriptions appears clearly manifested in the table. The CIPT and FOPT results have been finally averaged, but keeping conservatively the smaller errors because uncertainties are fully correlated.

\begin{table}[t]
\tbl{Determinations of $\alpha_{s}(m_{\tau}^{2})$ in the $V+A$ channel.\protect\cite{Pich:2016bdg}}
{\begin{tabular}{@{}cccc@{}} \toprule
Method  & \multicolumn{3}{c}{$\alpha_{s}(m_{\tau}^{2})$}
\\[1.5pt] \cline{2-4}
& \raisebox{-2pt}{CIPT} & \raisebox{-2pt}{FOPT} & \raisebox{-2pt}{Average}
\\ \colrule
ALEPH moments & $0.339 \,{}^{+\, 0.019}_{-\, 0.017}$ &
$0.319 \,{}^{+\, 0.017}_{-\, 0.015}$ & $0.329 \,{}^{+\, 0.017}_{-\, 0.015}$
\\[3pt]
Modified ALEPH moments  & $0.338 \,{}^{+\, 0.014}_{-\, 0.012}$ &
$0.319 \,{}^{+\, 0.013}_{-\, 0.010}$ & $0.329 \,{}^{+\, 0.013}_{-\, 0.010}$
\\[3pt]
$A^{(2,m)}$ moments  & $0.336 \,{}^{+\, 0.018}_{-\, 0.016}$ &
$0.317 \,{}^{+\, 0.015}_{-\, 0.013}$ & $0.326 \,{}^{+\, 0.015}_{-\, 0.013}$
\\[3pt]
$s_0$ dependence  & $0.335 \pm 0.014$ &
$0.323 \pm 0.012$ & $0.329 \pm 0.012$
\\[3pt]
Borel transform  & $0.328 \, {}^{+\, 0.014}_{-\, 0.013}$ &
$0.318 \, {}^{+\, 0.015}_{-\, 0.012}$ & $0.323 \, {}^{+\, 0.014}_{-\, 0.012}$
\\ \botrule
\end{tabular}}
\label{tab:summary}
\end{table}

The determination in the first line follows the same procedure adopted in the ALEPH analysis of Ref.~\refcite{Davier:2013sfa}. Taking $s_0=m_\tau^2$ and the weights
\begin{equation}\label{kl}
\omega_{kl}(s)\; =\;\left(1-\frac{s}{m_{\tau}^{2}}\right)^{2+k}\left( \frac{s}{m^{2}_{\tau}} \right)^{l}\left( 1+\frac{2s}{m_{\tau}^{2}}   \right) \, ,
\end{equation}
we have performed a global fit to the corresponding moments with
$(k,l)=\{(0,0), (1,0), (1,1), (1,2), (1,3)\}$. These weights incorporate the phase-space and spin-1 factors in Eq.~\eqn{eq:RtauSpectral}, allowing for a direct use of the measured hadronic distribution and the inclusion of the precise determination of $R_{\tau,V+A}$
with a universality-improved leptonic branching ratio (subtracting the small contribution of final states with non-zero strangeness). With these five moments, we have made a global fit of $\alpha_s(m_\tau^2)$, the gluon condensate, $\cO_6$ and $\cO_8$. To assess possible errors associated with neglected higher-order condensates (the highest moment involves power corrections with $D\le 16$), a second fit including $\cO_{10}$ has been performed and the variation on the fitted value of the strong coupling has been included as an additional uncertainty.
As expected, the extracted condensates have large relative errors exhibiting the very little sensitivity to power corrections, and a quite precise value of $\alpha_{s}(m_{\tau}^{2})$ is obtained. Our results are in very good agreement with Ref.~\refcite{Davier:2013sfa}, although our enlarged errors are more
conservative.

We have repeated the fits, taking away the factor $( 1+ 2s/m_{\tau}^2)$ from the weights \eqn{kl}. Although one loses the additional information from the $\tau$ lifetime and leptonic branching ratios, this eliminates the highest-dimensional condensate contribution to every moment. The fitted values, shown in the second line of Table~\ref{tab:summary}, are in perfect agreement with the results of the previous fit (first line) and are even more precise. This shows again the insensitivity to higher-order power corrections. Moreover, it suggests that our error estimates are perhaps too conservative.

A different strategy consists in using optimal weights which are only sensitive to specific condensate dimensions. Particularly suitable are the doubly-pinched weights
\bel{eq:omega-2n}
\omega^{(2,m)}(x)\; =\; (1-x)^2\;\,\sum_{k=0}^{m}\, (k+1)\, x^k \; =\;
1-(m + 2)\, x^{m + 1} + (m + 1)\, x^{m + 2}\, .
\ee
Their corresponding moments $A^{(n,m)}(s_0)\equiv A^{\omega^{(n,m)}}\! (s_0)$ only receive condensate contributions from $\cO_{2 (m+2)}$ and $\cO_{2 (m+3)}$. A combined fit of five different $A^{(2,m)}$ moments ($1\le m\le 5$) gives the results shown in the third line of Table~\ref{tab:summary}. We have made a global fit with four free parameters, assuming $\cO_{12}=\cO_{14}=\cO_{16}=0$. To account for these missing power corrections, the fit has been repeated with the inclusion of $\cO_{12}$ and the variation in the fitted value of $\alpha_{s}(m_{\tau}^{2})$ has been taken as an additional uncertainty. The agreement with the results obtained in the previous fits is excellent.

Similar results (not included in the table) are obtained from a global fit to four $A^{(n,0)}$ ($0\le n\le 3$) moments based on the n-pinched weights
\bel{eq:omega-n0}
\omega^{(n,0)}(x)\; =\; (1-x)^n\; =\;\sum_{k=0}^n \, (-1)^k\, \left( \ba{c} n \\ k \ea\right)\, x^k\, ,
\ee
which receive corrections from all condensates with $D\le 2(n+1)$, but are protected against duality violations for $n\not= 0$.

Neglecting all non-perturbative effects, one can determine $\alpha_{s}(m_{\tau}^{2})$ from a single moment. This interesting exercise has been also done in Ref.~\refcite{Pich:2016bdg}, making 13 separate extractions of the strong coupling with six $A^{(2,m)}$ moments ($0\le m\le 5$), six
$A^{(1,m)}$ moments ($0\le m\le 5$) based on the weights
\bel{eq:omega-1n}
\omega^{(1,m)}(x)\; =\; 1-x^{m+1}\; =\; (1-x)\;\,\sum_{k=0}^{m}\, x^k\, ,
\ee
which are only sensitive to $\cO_{2 (m+2)}$, and the moment $A^{(0,0)}$ where OPE corrections are absent but it is very exposed to duality-violation effects. In all cases, the resulting determinations of the strong coupling are in agreement with the values in Table~\ref{tab:summary}, reflecting the minor numerical role of the neglected non-perturbative corrections.

\section{Dependence on $s_0$}

\begin{figure}[tbh]
\centerline{
\includegraphics[width=0.495\textwidth]{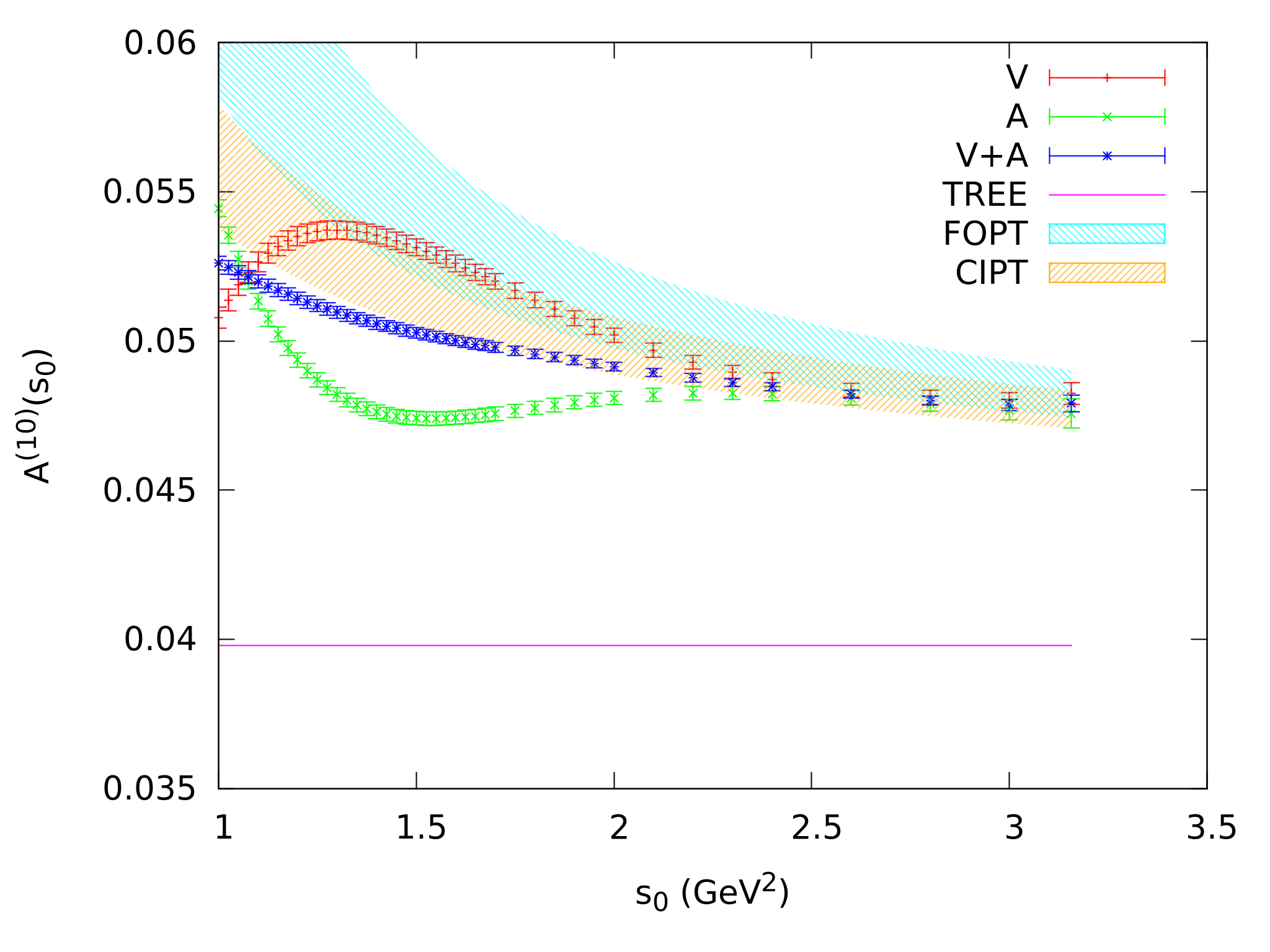}\hskip .2cm
\includegraphics[width=0.495\textwidth]{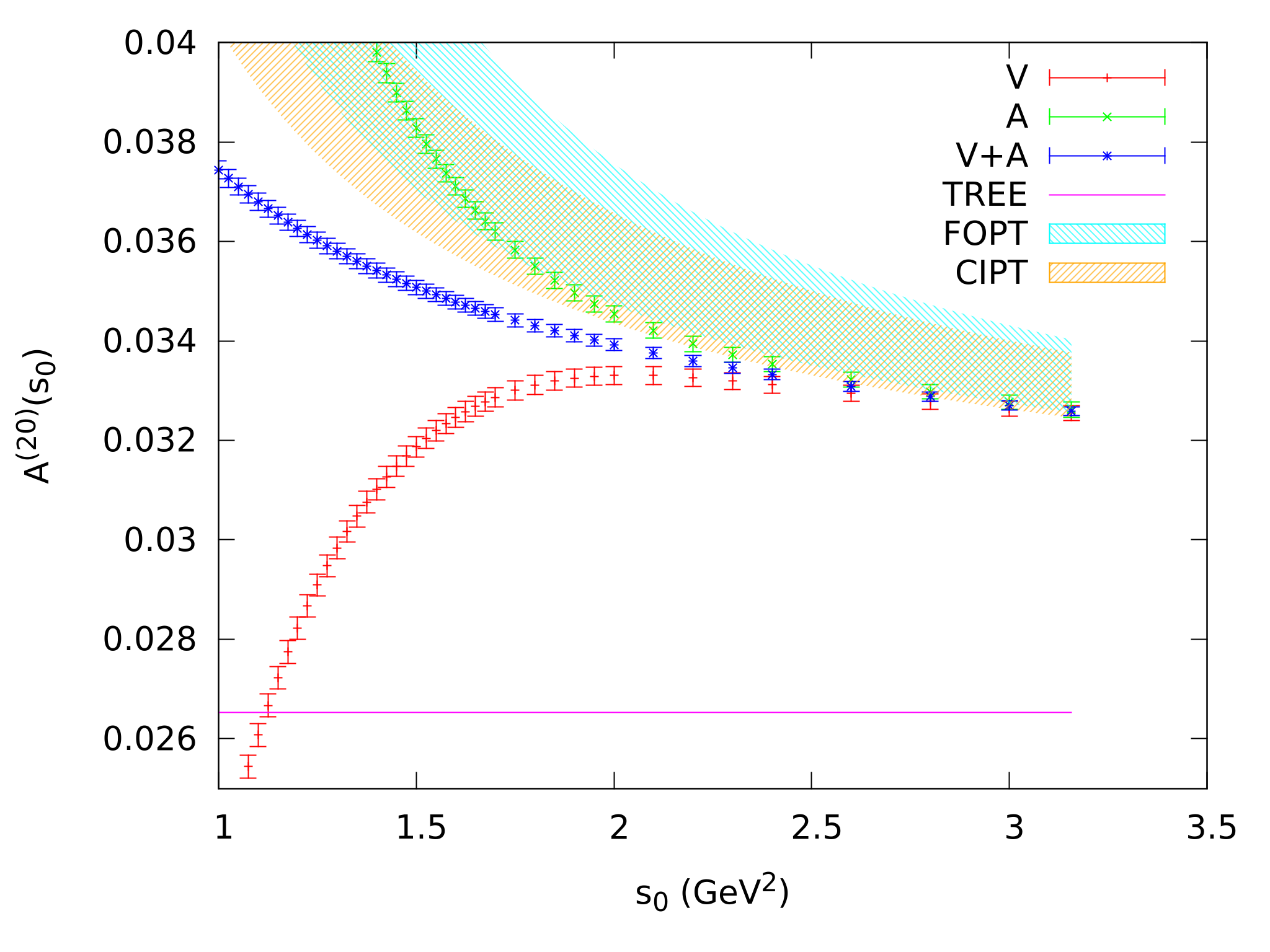}
}
\vspace*{8pt}
\caption{\label{pinchs}
Dependence on $s_0$ of the experimental moments $A^{(1,0)} (s_{0})$ (left) and $A^{(2,0)} (s_{0})$ (right), together with their purely CIPT and FOPT perturbative predictions for
$\alpha_s(m_\tau^2) = 0.329 \,{}^{+\, 0.017}_{-\, 0.015}$.
Data points are shown for the $V$ (red), $A$ (green) and $\frac{1}{2}\, (V+A)$ (blue)
channels.\protect\cite{Pich:2016bdg}}
\end{figure}

The non-perturbative contributions should be reflected in a distinctive $s_0$ dependence of the different moments. The power correction to the moment $A^{(1,m)}$ scales as $1/s_0^{m+2}$, while the $A^{(2,m)}$ moments get $1/s_0^{m+2}$ and $1/s_0^{m+3}$ corrections. Fig.~\ref{pinchs} shows the experimental moments $A^{(1,0)} (s_{0})$ and $A^{(2,0)} (s_{0})$ as function of $s_0$, in the $V$, $A$ and $\frac{1}{2}\, (V+A)$ channels, together with their predicted values with
$\alpha_s(m_\tau^2) = 0.329 \,{}^{+\, 0.017}_{-\, 0.015}$, neglecting all non-perturbative contributions.

The moment $A^{(1,0)} (s_{0})$, which can only get corrections from $\cO_4$, exhibits a surprisingly good agreement with its pure perturbative prediction in all channels ($V$, $A$ and $V+A$). In spite of being only protected by a single pinch factor, the data points closely follow the central values predicted by CIPT, above $s_0\sim 2\;\mathrm{GeV}^2$. In that energy range non-perturbative contributions appear to be too small to become numerically
visible within the much larger perturbative uncertainties covering the shades areas of the figure.
The splitting at lower values of $s_0$ of the $V$ and $A$ moments must be assigned to
duality violations, since their $D=4$ power corrections are approximately equal. However, these duality-violation effects clearly compensate in $V+A$, with an impressively flat distribution of the experimental data which does not deviate from the $1\sigma$ perturbative range even at $s_0 \sim 1\;\mathrm{GeV}$. A similar behaviour is observed for $A^{(0,0)} (s_{0})$, a moment without OPE corrections.

$A^{(2,0)} (s_{0})$ looks slightly more sensitive to non-perturbative contributions and seems to prefer a power correction with different signs for $V$ and $A$, which cancels to a good extend in $V+A$. This fits nicely with the expected $\cO_{6,V/A}$ contribution, although the merging of the $V$, $A$ and $V+A$  curves above $s_0\sim 2.2\;\mathrm{GeV}^2$ suggests a very tiny numerical effect from this source at high invariant masses.

\begin{figure}[tb]
\centerline{
\includegraphics[width=0.495\textwidth]{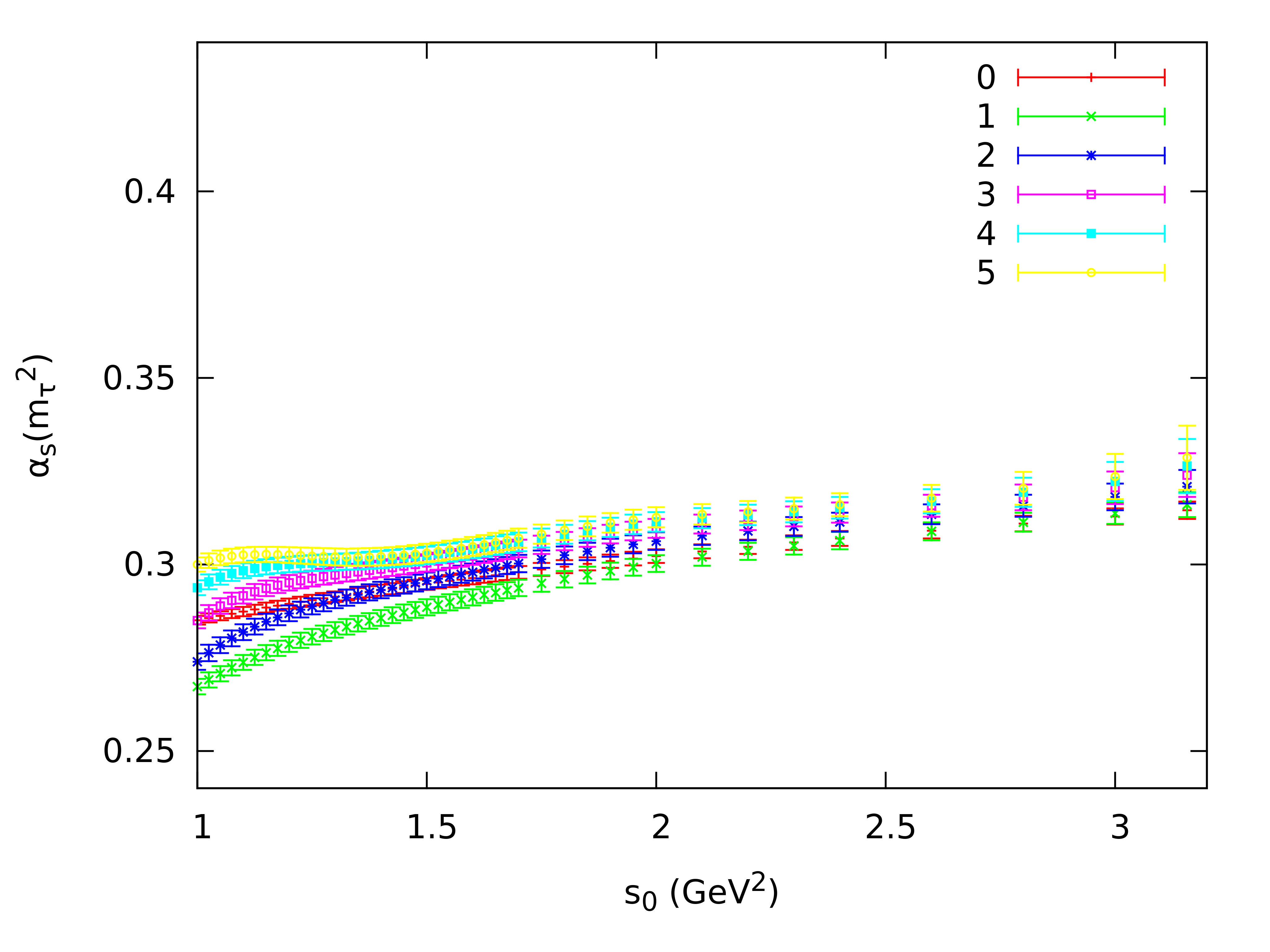}\hskip .2cm
\includegraphics[width=0.495\textwidth]{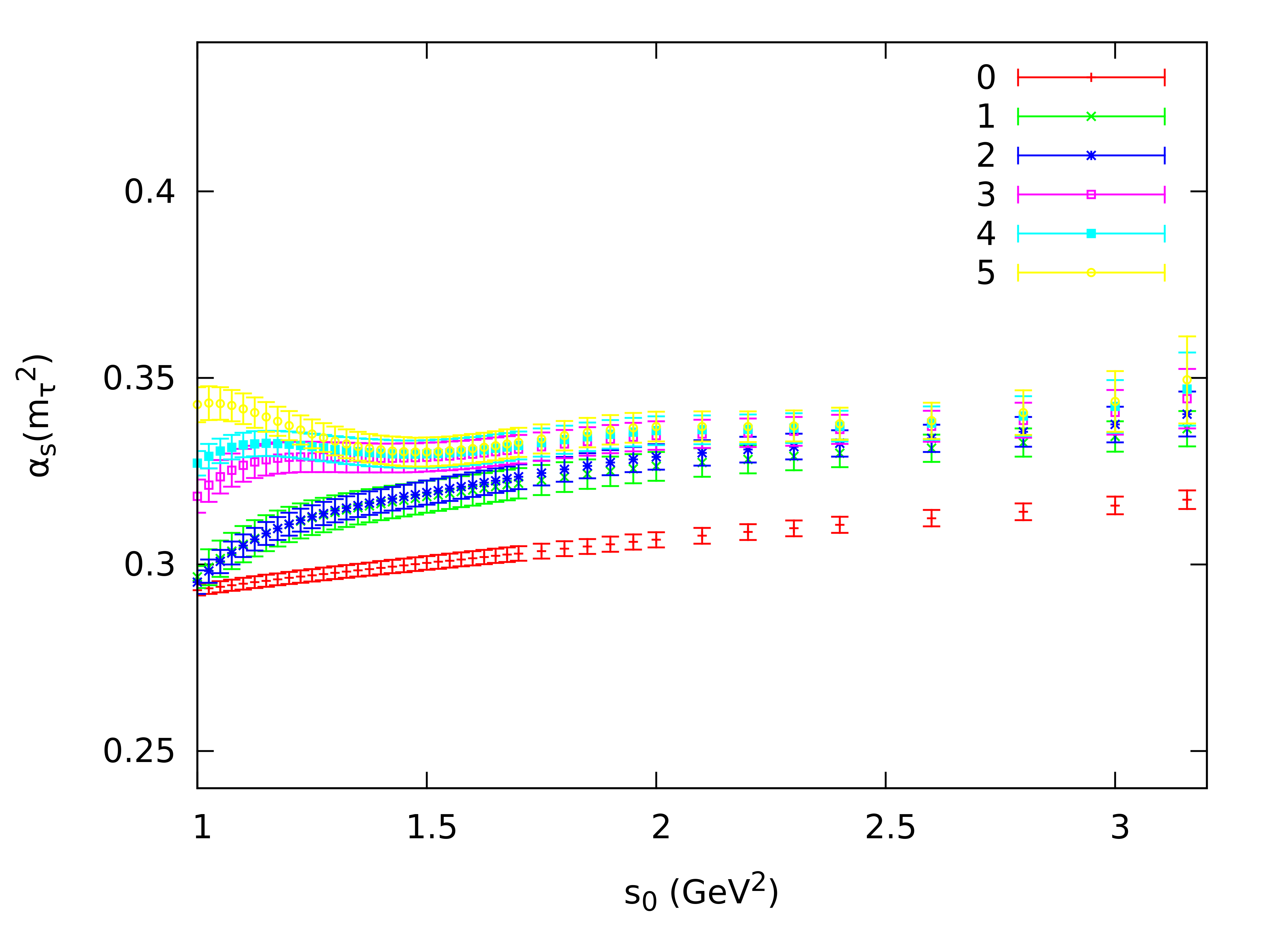}
}
\vspace*{8pt}
\caption{\label{todoa0}
$V+A$ determinations of $\alpha_{s}(m_{\tau}^{2})$ from different $A^{(2,m)} (s_{0})$ moments with $\{ m=0,...,5\}$, as function of $s_0$, fitted ignoring all non-perturbative corrections. FOPT fits are on the left and CIPT on the right. Only experimental uncertainties have been included.\protect\cite{Pich:2016bdg}}
\end{figure}

Fig.~\ref{todoa0} shows, as function of $s_0$, independent determinations of $\alpha_s(m_\tau^2)$ extracted from six different $A^{(2,m)} (s_{0})$ moments of the $V+A$ distribution, ignoring all non-perturbative effects. Very similar plots with seven $A^{(1,m)} (s_{0})$ moments can be found in Ref.~\refcite{Pich:2016bdg}. These moments get completely different non-perturbative corrections, carrying a broad variety of inverse powers of $s_0$. The clustering of all curves, exhibiting a similar functional dependence on $s_0$, strongly indicates that inverse power corrections are small for $V+A$. Notice that only the experimental errors are displayed in the figure. The small fluctuations of the different curves are well within the much larger perturbative uncertainties shown in Fig.\ref{pinchs}.

Fitting the $s_0$ dependence of a single $A^{(2,m)} (s_{0})$ moment of the $V+A$ distribution, one can determine the values of the strong coupling and the power corrections $\cO_{2(m+2)}$ and $\cO_{2(m+3)}$. Ref.~\refcite{Pich:2016bdg} has analyzed the three lowest moments with $m=0,1,2$, using the nine energy bins above $s_0 = 2.0\;\mathrm{GeV}^2$. For each moment, six different fits have been performed, varying the number of bins included in the fit between four and nine (always the highest-energy ones). The sensitivity to power corrections is very bad, as expected, but one finds an amazing stability in the extracted values of $\alpha_s(m_\tau^2)$. Including the information from the three moments and the nine energy bins, and adding as an additional uncertainty the small fluctuations observed when changing the number of fitted bins, one obtains the values of $\alpha_s(m_\tau^2)$ quoted in the fourth line of Table~\ref{tab:summary}. The excellent agreement with the more solid determinations in the first three lines of the table is quite unexpected, since now we are much more sensitive to violations of quark-hadron duality. In fact, as explicitly demonstrated in Ref.~\refcite{Pich:2016bdg}, when fitting the $s_0$ dependence of several consecutive bins one is using information about the local structure of the spectral function. However, the very flat shape of the $V+A$ hadronic distribution above $s_0 = 2.0\;\mathrm{GeV}^2$ implies small duality-violation effects in that region which, moreover, are very efficiently suppressed in the doubly-pinched moments $A^{(2,m)} (s_{0})$.

\section{Borel Transform}

When violations of duality are more important than power corrections, one can try to reduce them through the use of exponentially-suppressed moments of the type\cite{Pich:2016bdg}
\be
\omega_{a}^{(1,m)}(x)\; =\; \left(1- x^{m+1}\right)\; \mathrm{e}^{-ax} \, .
\ee
The exponential factor nullifies the highest invariant-mass region, but paying the price that all condensates contribute to every moment. For $a=0$ one recovers the $A^{(1,m)}(s_0)$ moments, only affected by $\cO_{2 (m+2)}$, while for $a\gg 1$ the OPE corrections become independent of $m$. Thus, if one determines with these moments the strong coupling, neglecting all non-perturbative contributions, the OPE corrections should manifest in a larger instability under variations of $s_0$ than in the $a=0$ case. The splitting among different moments, at a given value of $s_0$, should increase with a non-zero Borel parameter $a$, before they converge at $a\to\infty$.

\begin{figure}[tb]
\centerline{
\includegraphics[width=0.35\textwidth]{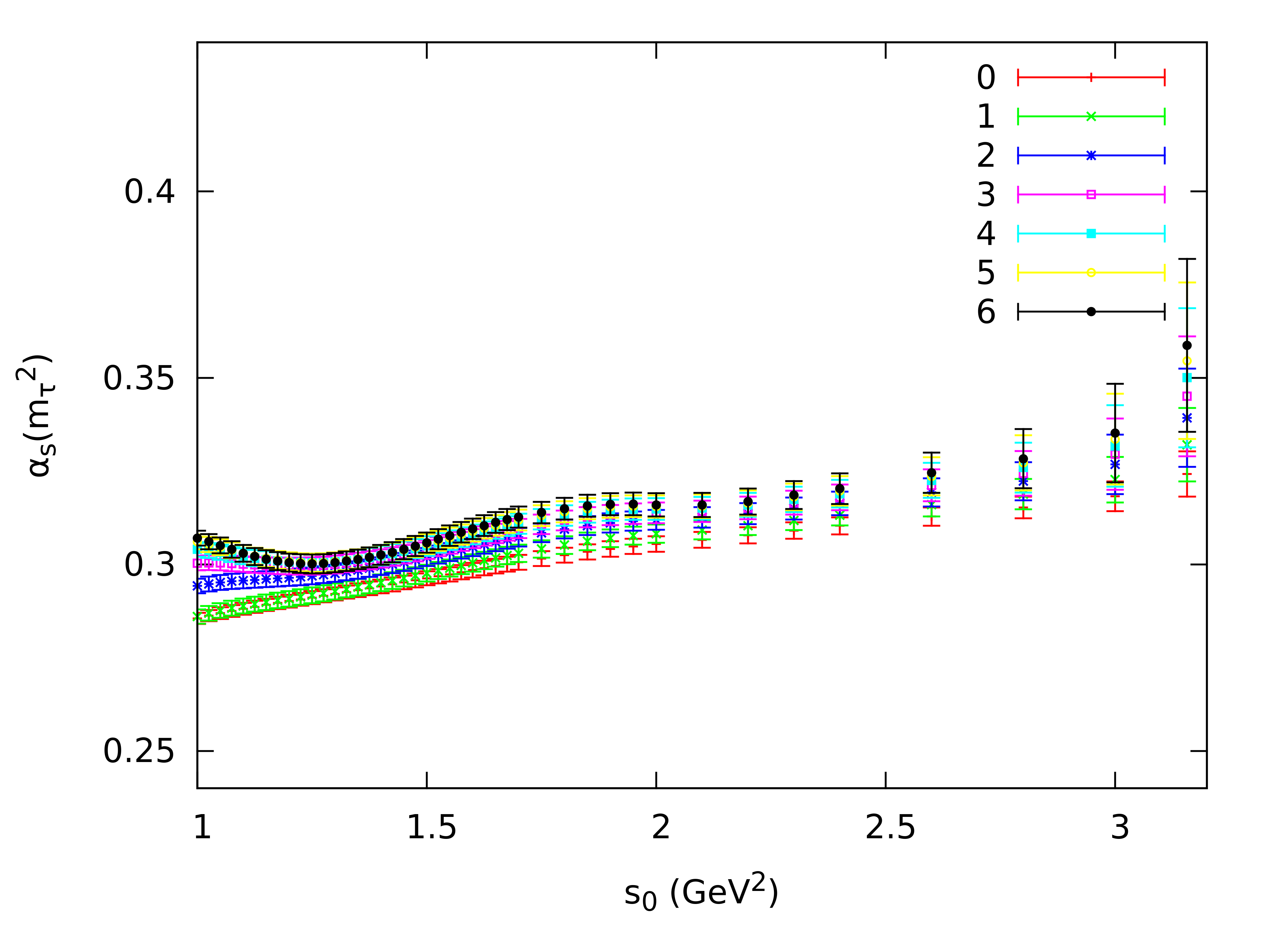}\hskip -.1cm
\includegraphics[width=0.35\textwidth]{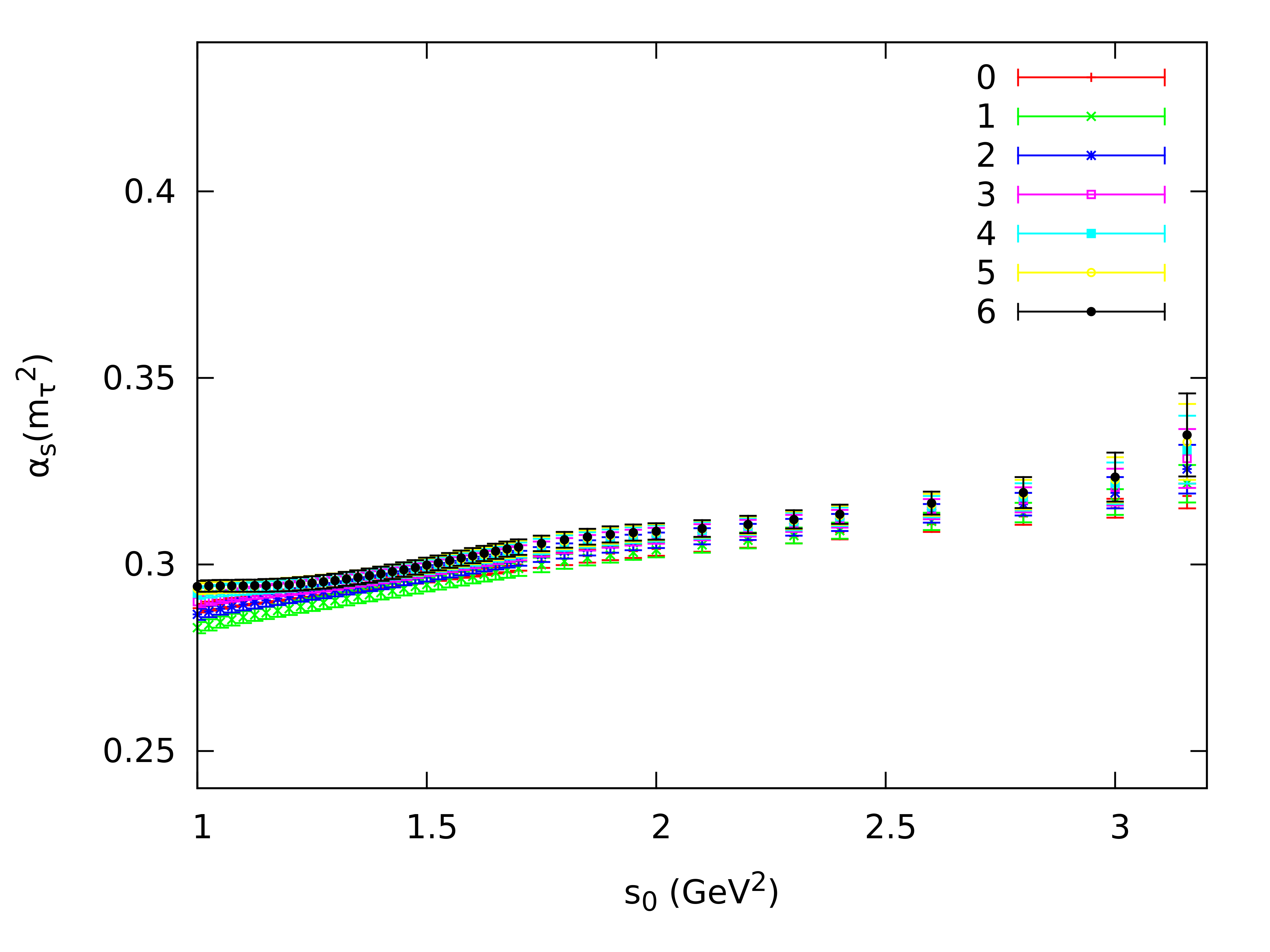}\hskip -.1cm
\includegraphics[width=0.35\textwidth]{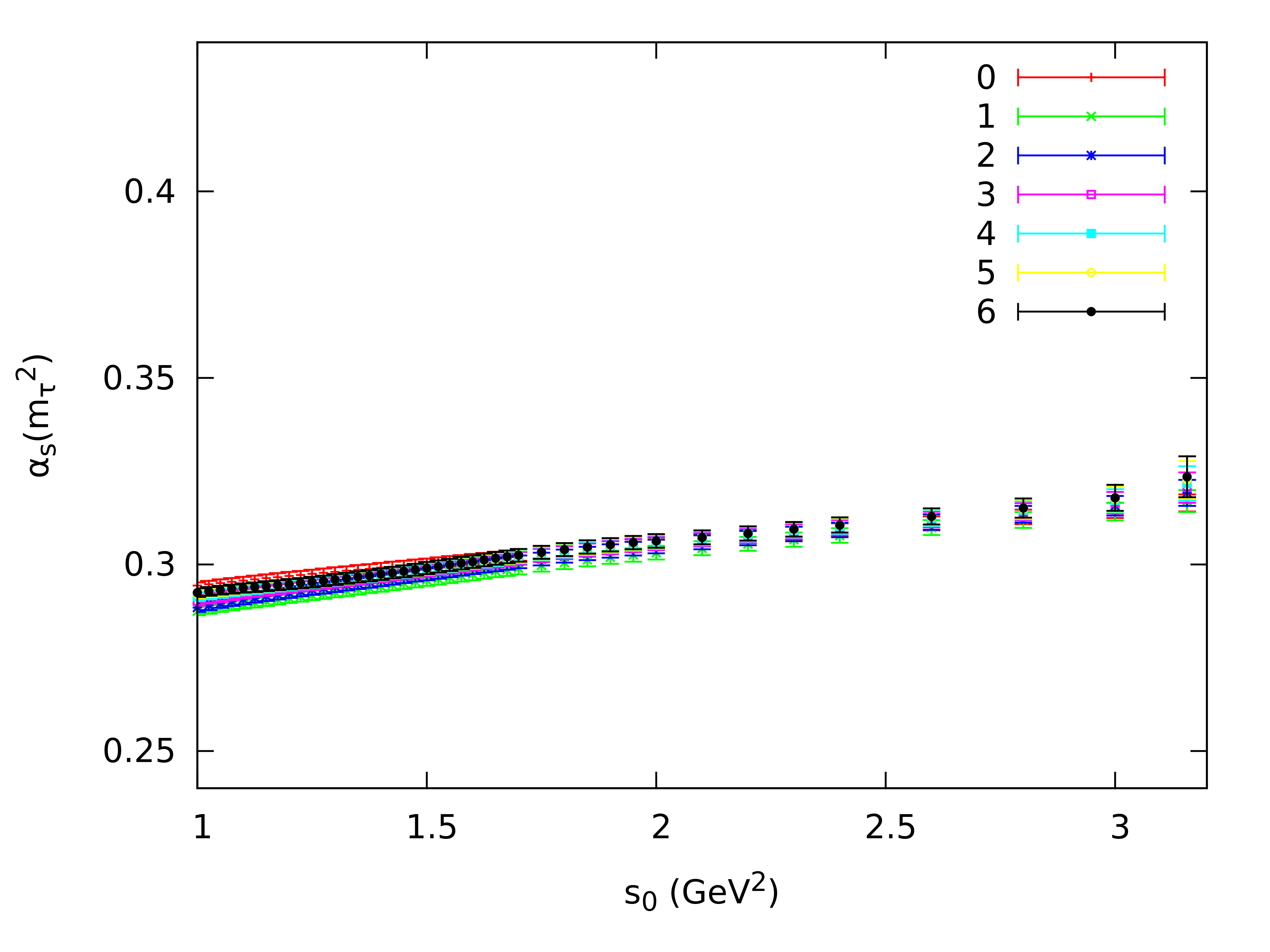}
}
\vspace*{8pt}
\caption{\label{transf2} FOPT determinations of $\alpha_{s}(m_{\tau}^{2})$ from the moments
$A_{V+A}^{\omega^{(1,m)}_a}(s_0)$, as function of $s_0$, ignoring all non-perturbative corrections and evaluated at $a=0$ (left), 1 (center) and 2 (right). Only experimental uncertainties have been included.\protect\cite{Pich:2016bdg}}
\end{figure}

\begin{figure}[tb]
\centerline{
\includegraphics[width=0.49\textwidth]{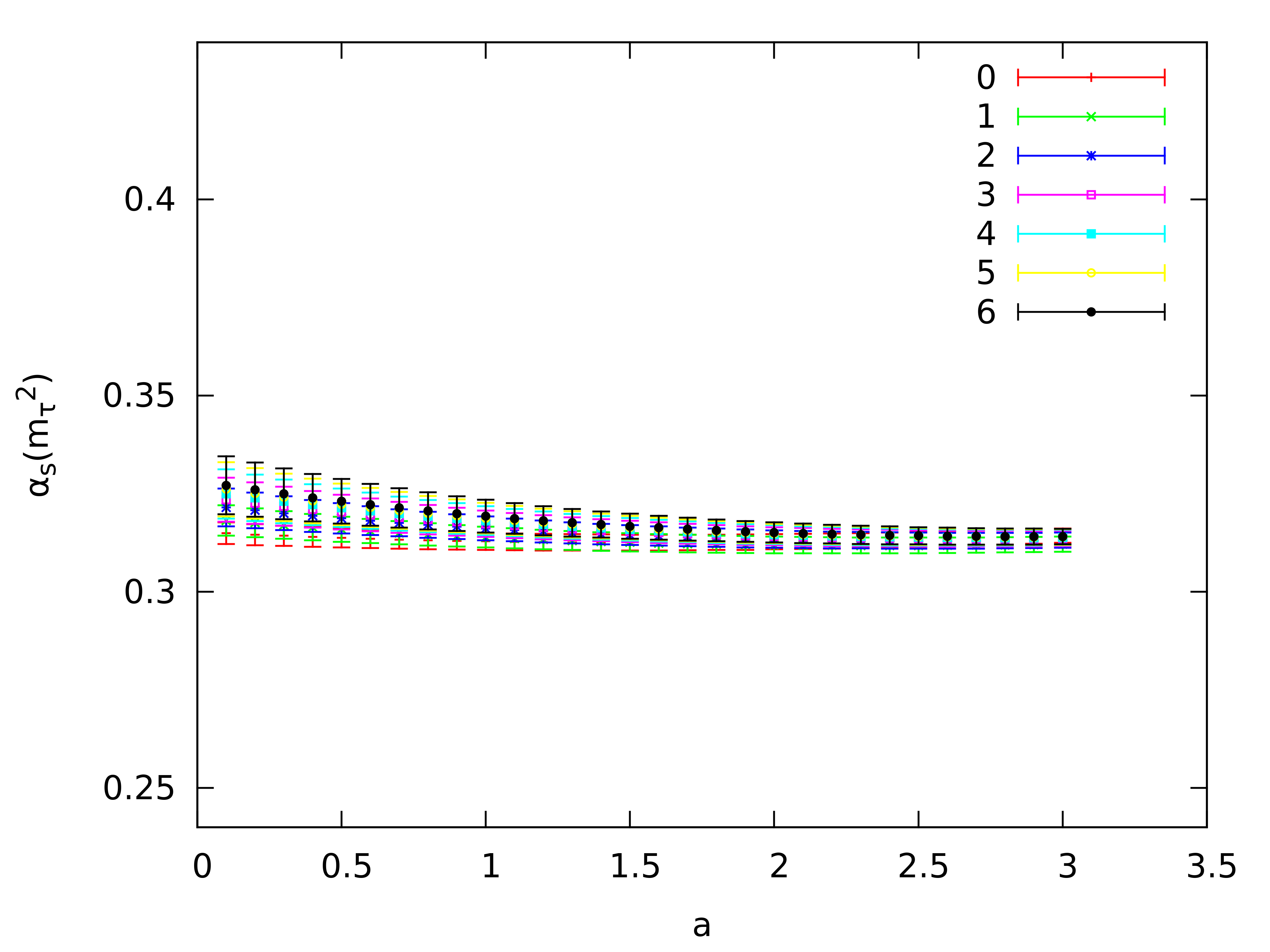}
}
\vspace*{8pt}
\caption{\label{bor2} FOPT determinations of $\alpha_{s}(m_{\tau}^{2})$ from the moments
$A_{V+A}^{\omega^{(1,n)}_a}(s_0)$, at $s_{0}=2.8\;\mathrm{GeV}^{2}$ and ignoring all non-perturbative corrections, for different values of $a$.
Only experimental uncertainties are included.\protect\cite{Pich:2016bdg}}
\end{figure}

The FOPT determinations of $\alpha_{s}(m_{\tau}^{2})$ extracted from seven $A_{V+A}^{\omega^{(1,m)}_a}(s_0)$ moments ($m=0,\cdots,6$), neglecting all non-perturbative contributions, are shown in Fig.~\ref{transf2} as function of $s_0$, for three different values of $a=0,1,2$. Clearly, one gets even more stable results when $a\not= 0$, and the different moments converge very soon when $a$ increases. For these weights, power corrections do not seem to be relevant in the plotted ranges of $a$ and $s_0$. Taking $s_0= 2.8\;\mathrm{GeV}^2$, the FOPT determinations of
$\alpha_{s}(m_{\tau}^{2})$ are plotted in Fig.~\ref{bor2} as function of $a$. We observe the existence of similar stability ranges in the two variables $s_0$ and $a$. Accepting for each moment all values of $\alpha_{s}(m_{\tau}^{2})$ in the Borel-stable region, including the information from all moments, and adding as additional theoretical uncertainties the differences among moments and the variations in the region $s_0\in [2,2.8]\;\mathrm{GeV}^2$, one gets the determination of $\alpha_{s}(m_{\tau}^{2})$ shown in the fifth line of Table~\ref{tab:summary}.

\begin{figure}[tb]
\centerline{
\includegraphics[width=0.35\textwidth]{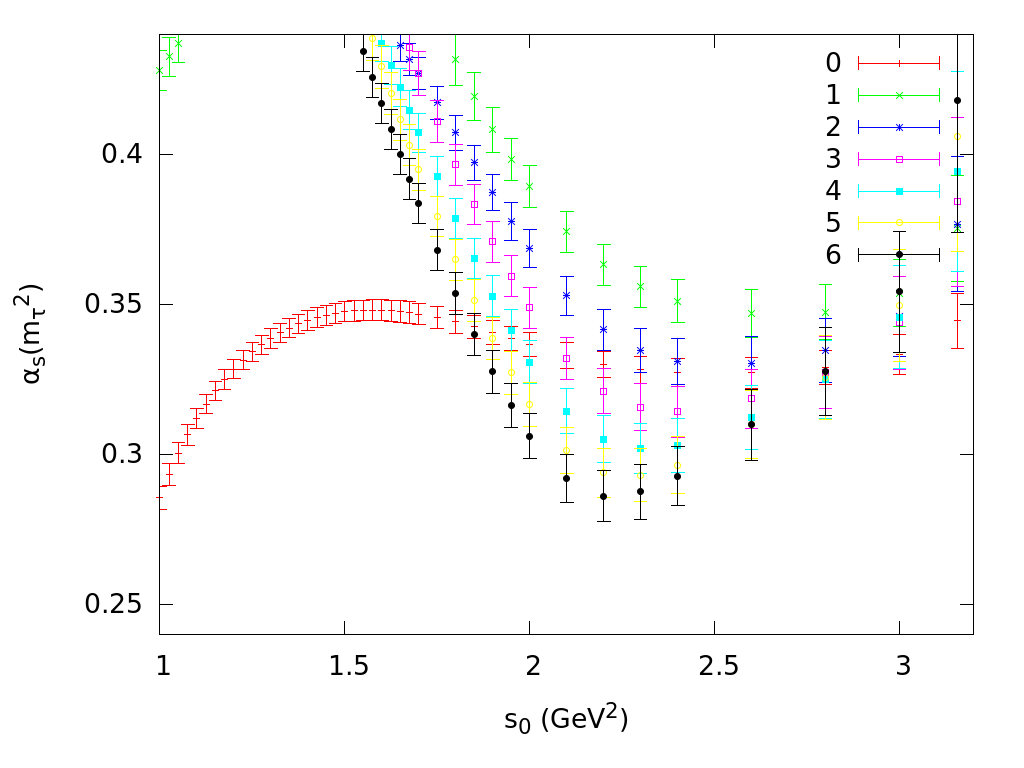}\hskip -.1cm
\includegraphics[width=0.35\textwidth]{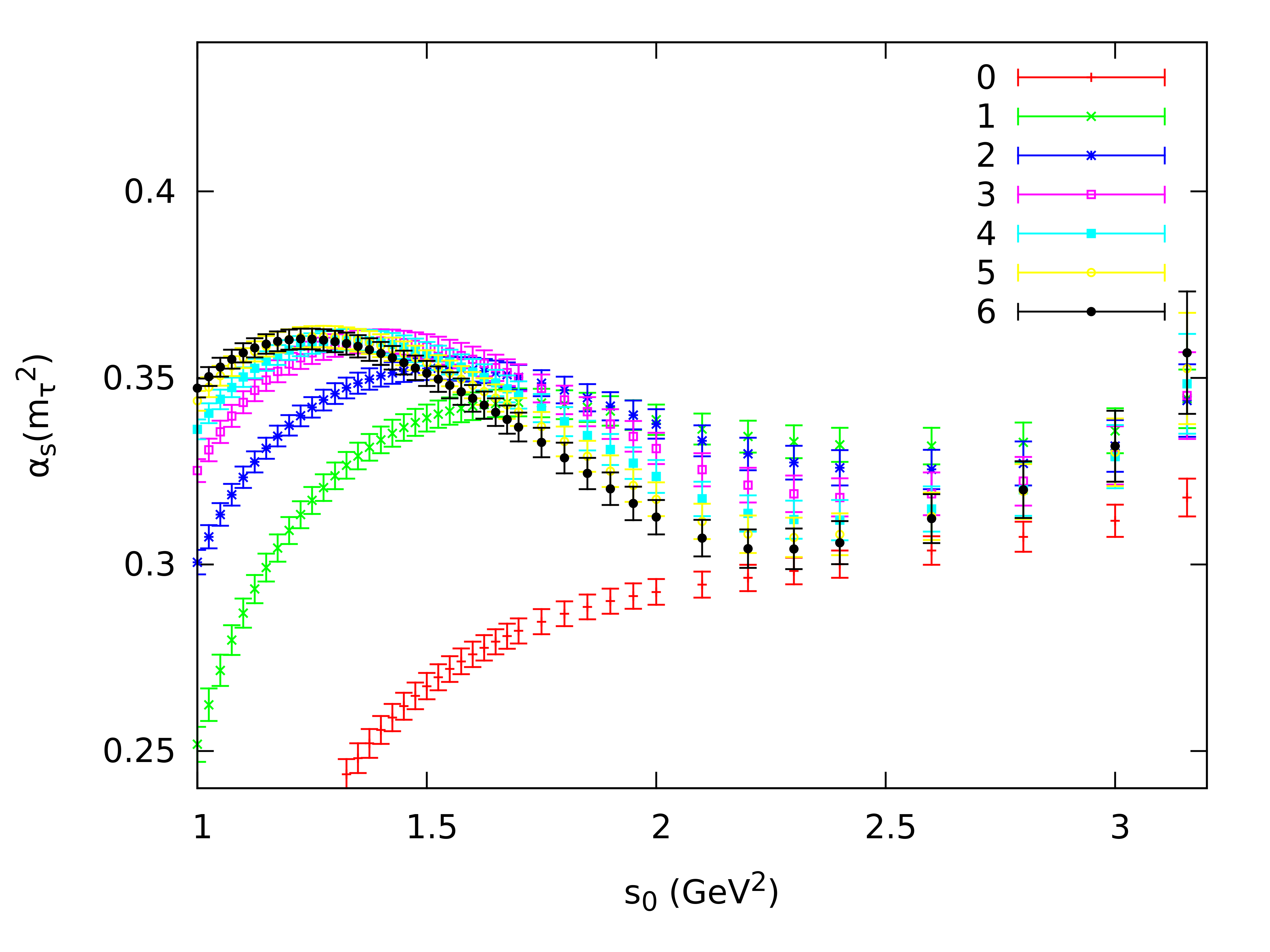}\hskip -.1cm
\includegraphics[width=0.35\textwidth]{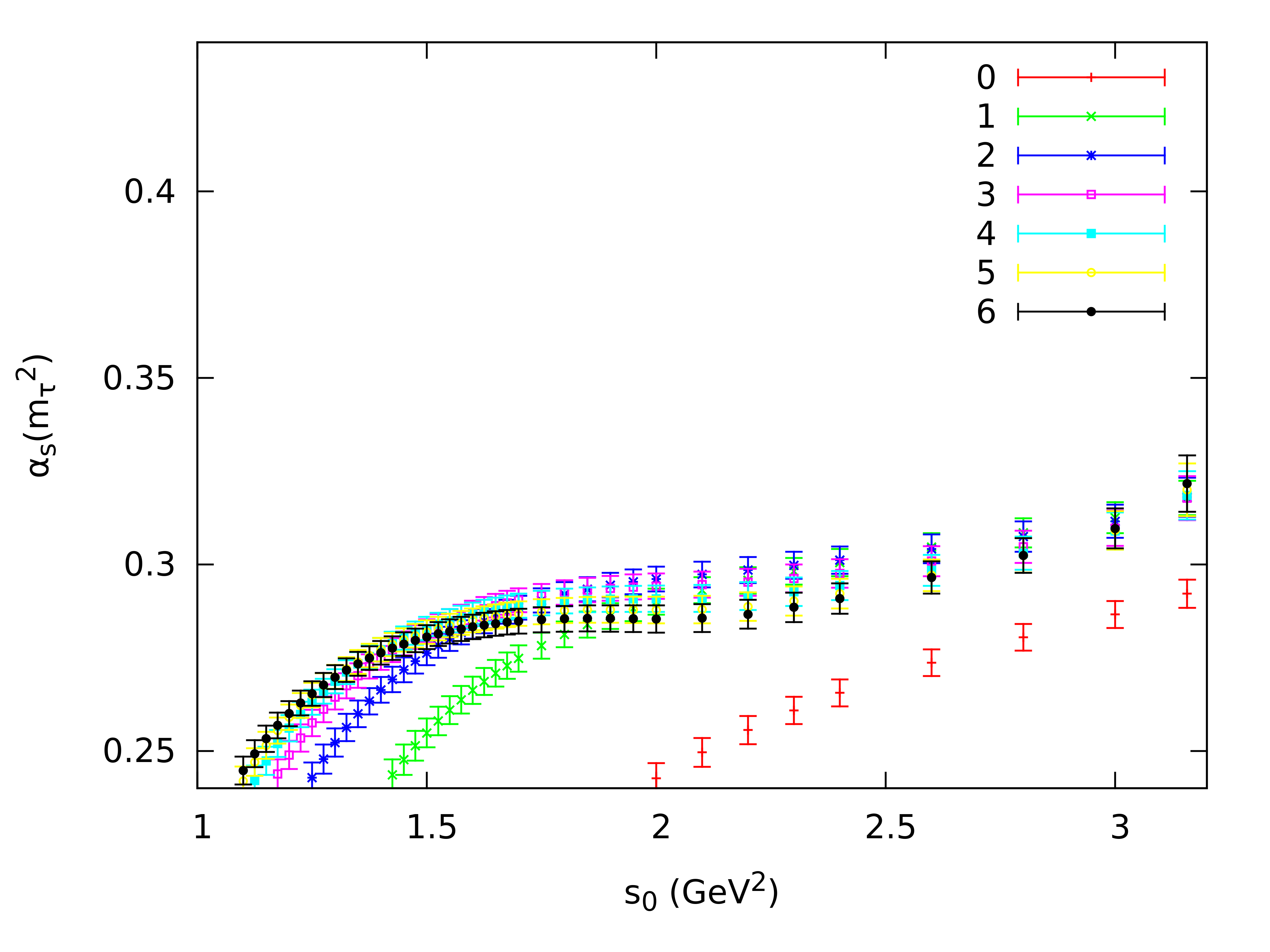}}
\vspace*{0.05cm}
\centerline{
\includegraphics[width=0.35\textwidth]{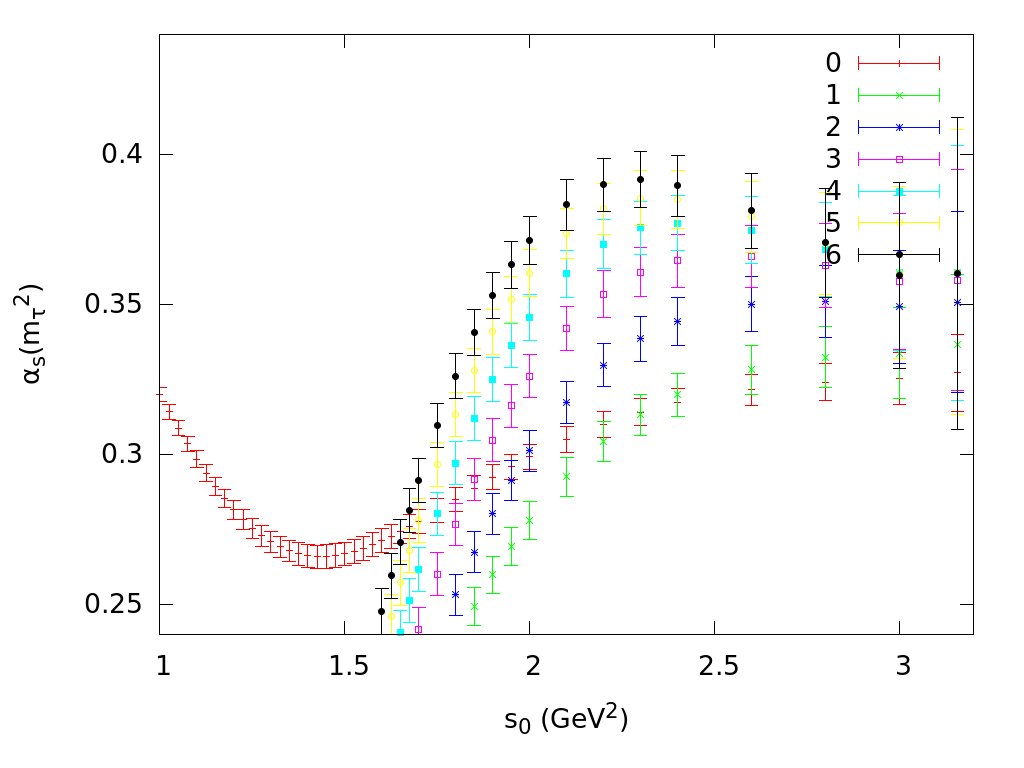}\hskip -.1cm
\includegraphics[width=0.35\textwidth]{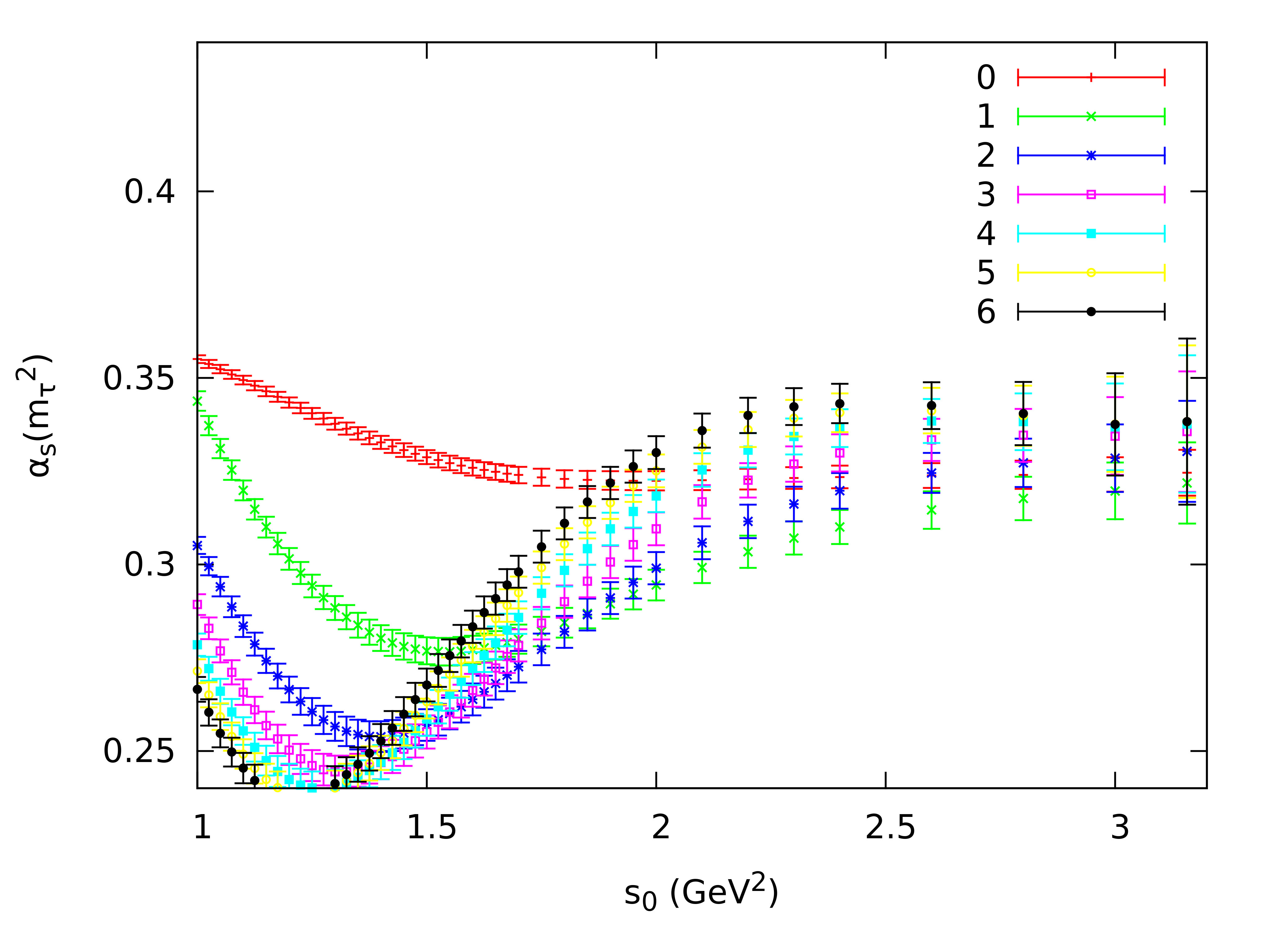}\hskip -.1cm
\includegraphics[width=0.35\textwidth]{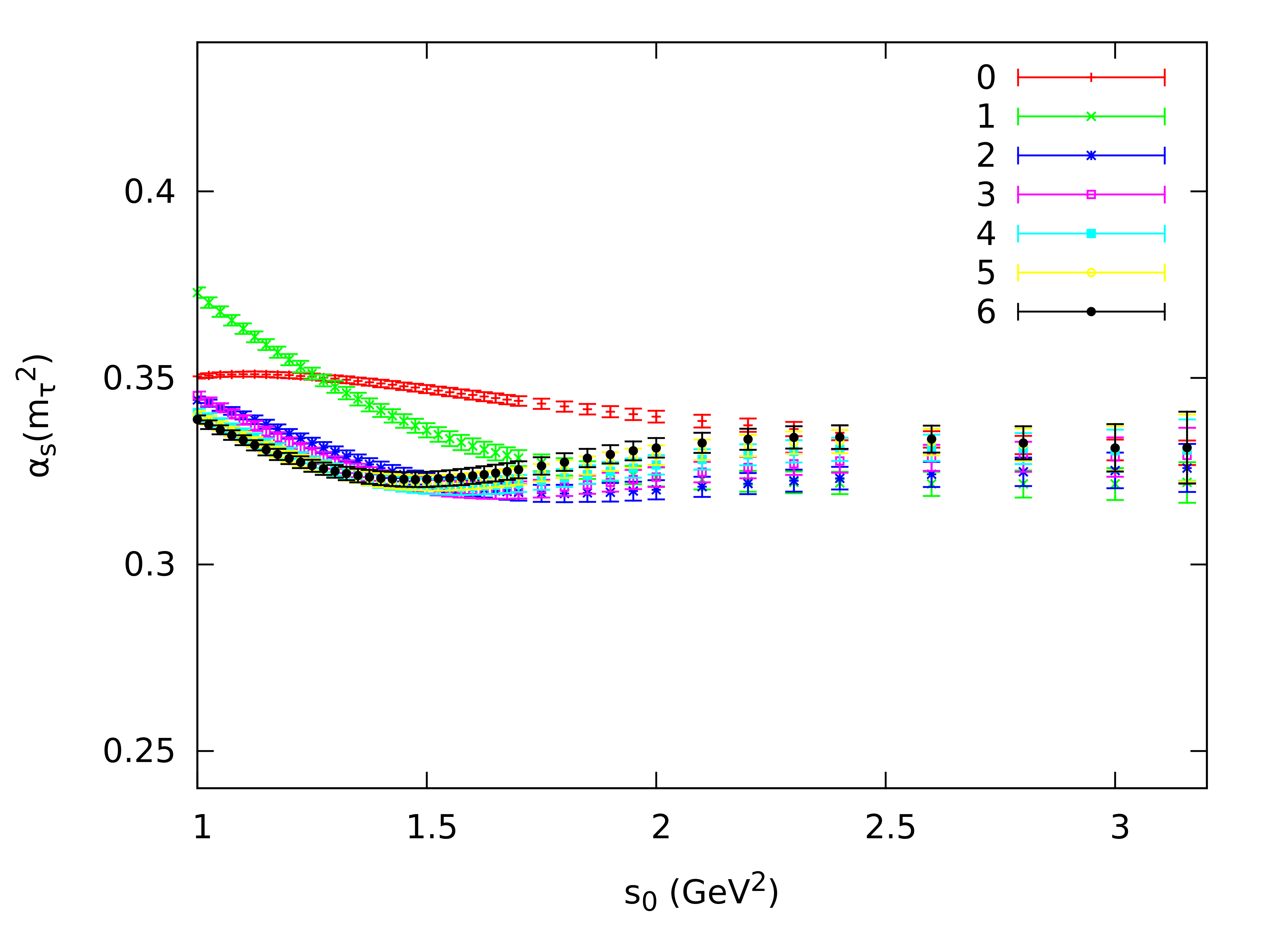}}
\vspace*{8pt}
\caption{\label{transf} CIPT determinations of $\alpha_{s}(m_{\tau}^{2})$ from $A_{V/A}^{\omega^{(1,m)}_a}(s_0)$ moments, as function of $s_0$ and ignoring all non-perturbative corrections, at $a=0$ (left), 1 (center) and 2 (right). The top (bottom) panels correspond to the $V$ ($A$) distribution. Only experimental uncertainties have been included.\protect\cite{Pich:2016bdg}}
\end{figure}

While this Borel determination of $\alpha_{s}(m_{\tau}^{2})$ agrees well with the previous ones, using quite different methods, it does not bring clear improvements because duality-violation effects are very suppressed in the $V+A$ case. The situation is different in the separate $V$ and $A$ analyses where violations of duality are more prominent at low and intermediate
invariant masses (see Figs.~\ref{fig:ALEPHsf} and \ref{pinchs}). With non-zero values of the Borel parameter $a$, one gets in the two channels nice stability regions both in $a$ and $s_0$, shown in Fig.~\ref{transf}, from which it is possible to derive precise determinations of $\alpha_{s}(m_{\tau}^{2})$ in a quite straightforward way:\cite{Pich:2016bdg}
\begin{align}
\alpha_{s}(m_\tau^2)^{V, \mathrm{CIPT}}\, & =\;  0.326 \, {}^{+0.021}_{-0.019}\, ,
\qquad\qquad
\alpha_{s}(m_\tau^2)^{A, \mathrm{CIPT}}&\hskip -.5cm =\; 0.325 \, {}^{+0.018}_{-0.014}\, ,
\no\\[5pt]
\alpha_{s}(m_\tau^2)^{V, \mathrm{FOPT}}& =\; 0.314 \, {}^{+0.015}_{-0.011}\, ,
\qquad\qquad
\alpha_{s}(m_\tau^2)^{A, \mathrm{FOPT}}&\hskip -.5cm  =\; 0.320 \, {}^{+0.019}_{-0.016}\, .
\end{align}
Thus, one finds an excellent consistency between the vector and axial-vector determinations which are, moreover, in good agreement with the $V+A$ results.

\section{Models of Duality Violation}

Instead of using clean moments where violations of duality are very suppressed, some recent works have focussed on observables much more sensitive to these uncontrollable effects,\cite{Boito:2014sta} 
modeling them through an ansatz that parametrizes the differences $\Delta\rho^{\mathrm{DV}}_{V/A}(s)$ between the physical spectral functions and their OPE approximations.\cite{Cata:2008ye} The model parameters are fitted to the experimental data and used to estimate the duality-violation correction to a given moment with the identity\cite{Cata:2008ye,Chibisov:1996wf,GonzalezAlonso:2010rn,Rodriguez-Sanchez:2016jvw}
\bel{eq:DVcorr}
\frac{i}{2}\;\oint_{|s|=s_{0}}
\frac{ds}{s_{0}}\;\omega(s)\,\left\{\Pi^{\phantom{\mathrm{OPE}}}_{V/A}(s) - \Pi^{\mathrm{OPE}}_{V/A}(s)\right\}
\; =\; -\pi\;\int^{\infty}_{s_{0}} \frac{ds}{s_{0}}\;\omega(s)\; \Delta\rho^{\mathrm{DV}}_{V/A}(s)
\, .
\ee

Let us consider the slightly generalized ansatz (in GeV units)
\bel{eq:DVparam}
\Delta\rho^{\mathrm{DV}}_{V/A}(s)\; =\; s^{\lambda_{V/A}}\; e^{-(\delta_{V/A}+\gamma_{V/A}s)}\;\sin{(\alpha_{V/A}+\beta_{V/A}s)}\, ,
\qquad\qquad
s> \hat s_0\, ,
\ee
which for $\lambda_{V/A} =0$ coincides with the model assumed in Refs.~\refcite{Boito:2014sta}.
The combination of a dumping exponential with an oscillatory function is expected to reasonably describe the fall-off of duality violations at very high energies, but this functional form is completely ad-hoc and difficult to justify at low energies.
Assuming the ansatz to be valid above $\hat s_0\sim 1.5\;\mathrm{GeV}^2$, Refs.~\refcite{Boito:2014sta} advocate to extract $\alpha_s(m_\tau^2)$, the vacuum condensates and the ansatz parameters from a global fit to the $s_0$ dependence of $A^{\omega}_{V/A}(s_{0})$ moments.
Since there are far too many parameters to be fitted to a highly-correlated data set, they concentrate in the moment $A^{(0,0)}_{V/A}(s_{0})$ which is very exposed to duality-violation effects and does not receive OPE corrections.

The problem with this strategy is that a fit with $n$ $s_0$ points of the $A^{(0,0)}_{V/A}(s_{0})$ moment is equivalent to a fit of
\be
\left\{ A^{(0,0)}_{V/A}(s_{0})\, ,\, \rho_{V/A}^{\phantom{()}}(s_{0})\, ,\, \rho_{V/A}^{\phantom{()}}(s_{0}+\Delta s_{0})\, ,\,\cdots\, ,\, \rho_{V/A}^{\phantom{()}}(s_{0}+(n-2)\Delta s_{0})\right\}\, .
\ee
Thus, $n-1$ points are dedicated to perform a direct fit of the spectral function.
Since the OPE is not valid in the real axis, one loses theoretical control and gets at best an effective model description with unclear relation to QCD.

\begin{figure}[tb]
\centerline{
\includegraphics[width=0.49\textwidth]{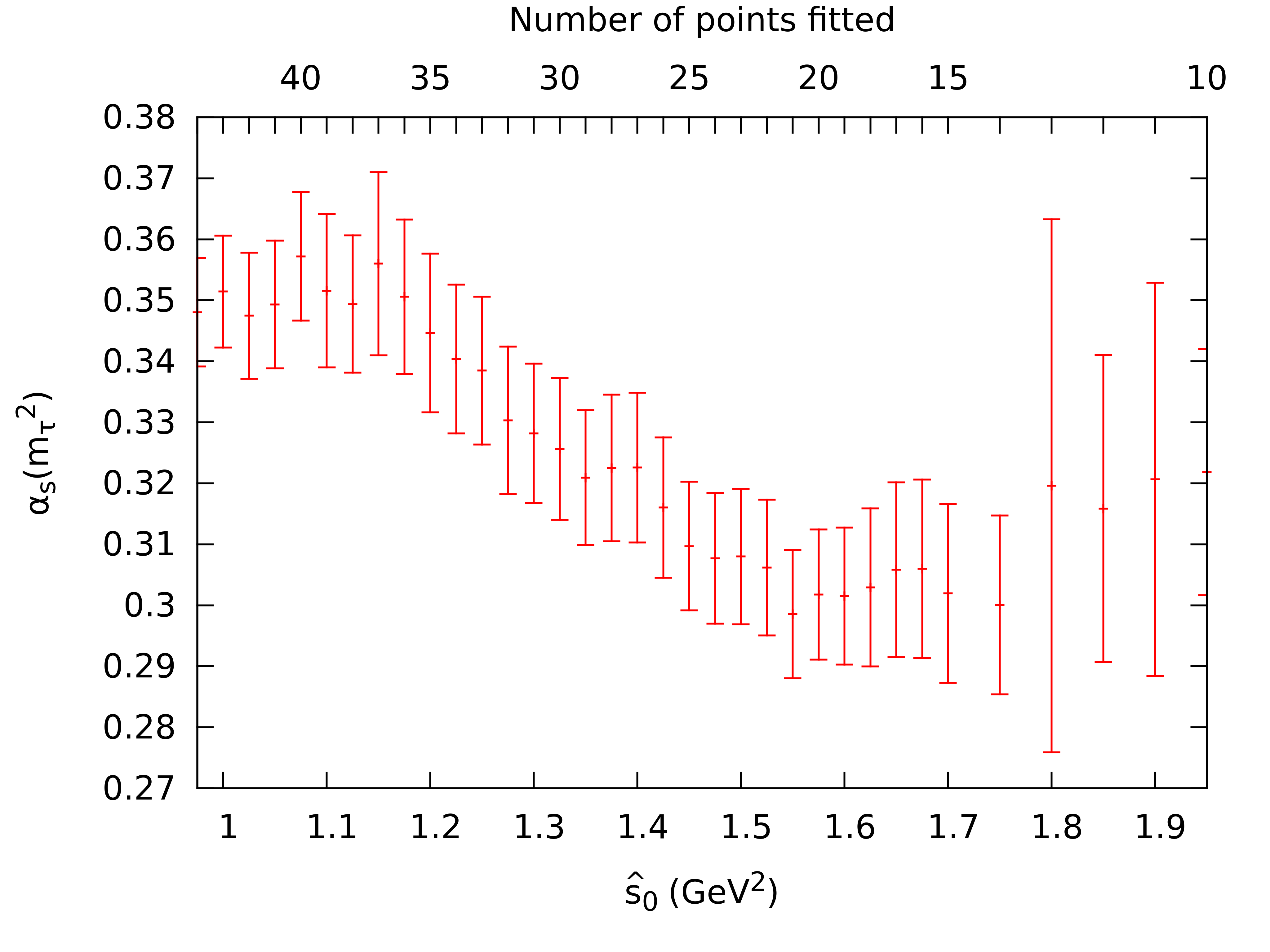}
\includegraphics[width=0.49\textwidth]{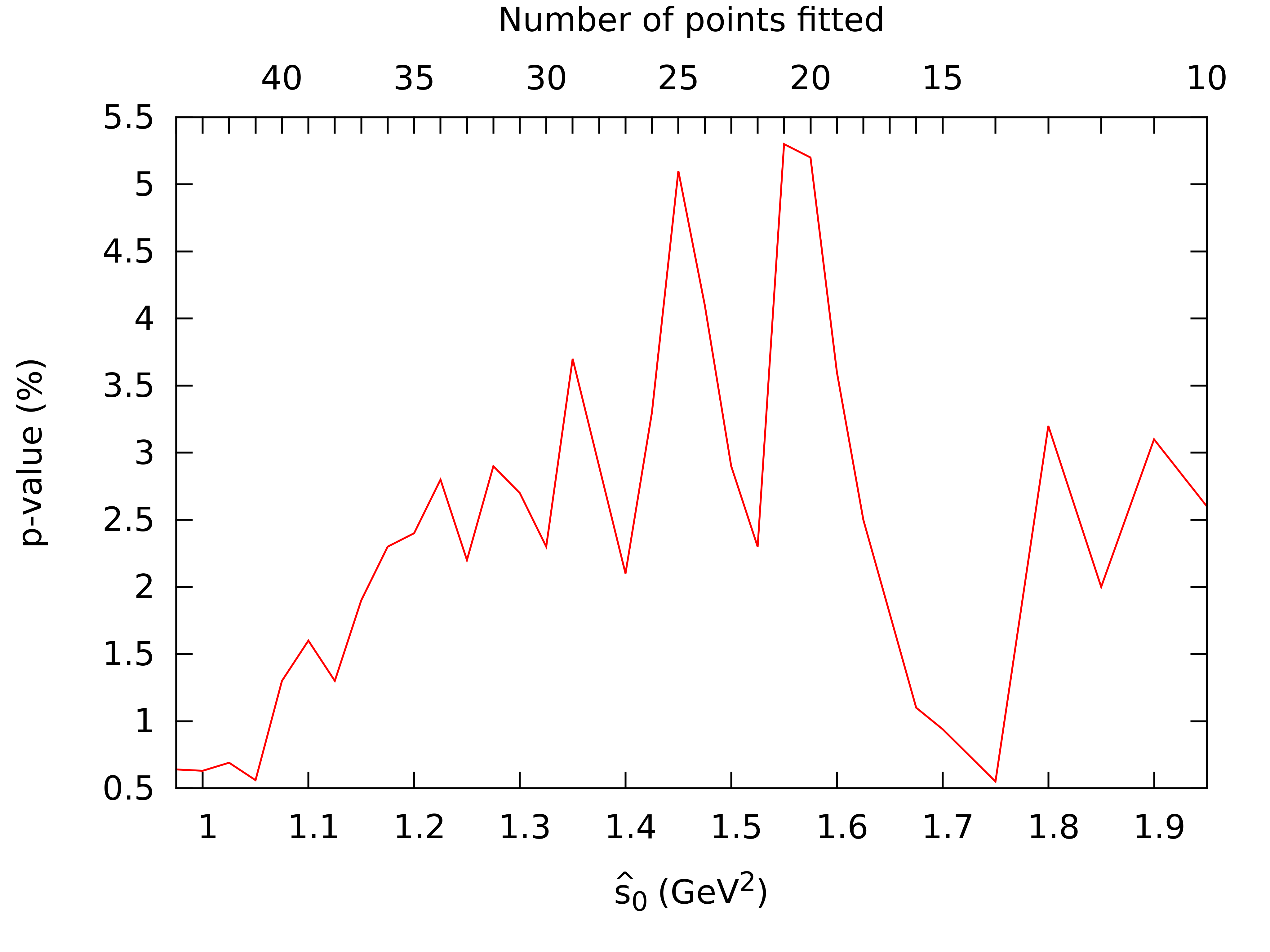}
}
\vspace*{8pt}
\caption{\label{boitofig} FOPT determination of $\alpha_s(m_\tau^2)$ from the $s_0$ dependence of $A^{(0 0)}_{V}(s_0)$, fitting all $s_0$ bins with $s_{0}>\hat s_0$, as function of $\hat s_0$, using the approach of Refs. \protect\refcite{Boito:2014sta}.}
\end{figure}

The axial channel is not useful to extract $\alpha_s$ with this method because the tail of the $a_1(3\pi)$ resonance extends to quite large values of $s$, and the energy bins where the ansatz
could be justified have too large experimental errors.
Taking $\lambda_{V/A} =0$, we have reproduced the results of Refs.~\refcite{Boito:2014sta}, extracted from $A^{(0,0)}_{V}(s_{0})$. They are displayed in Fig.~\ref{boitofig} that shows, as function of $\hat s_0$, the value of $\alpha_s(m_\tau^2)$ obtained from a fit to all energy bins above $\hat s_0$, using FOPT (similar results are obtained with CIPT). The right panel gives the p-values of the different  fits, making evident their very poor statistical quality for all $\hat s_0$ values. Ref.~\refcite{Boito:2014sta} chooses to perform the $\alpha_s(m_\tau^2)$ determination at $\hat s_0= 1.55\;\mathrm{GeV}^2$ because it has the larger (but still too small) p-value. However, this procedure does not have any good justification. The p-value falls dramatically when one moves from this magic point, becoming worse at higher $\hat s_0$ values where the model should work better. Moreover, the extracted value of $\alpha_s(m_\tau^2)$ is very unstable. Just removing from the fit one of the 20 fitted points, one observes fluctuations of the order of $1\sigma$.

\begin{figure}[t]
\centerline{\includegraphics[width=0.5\textwidth]{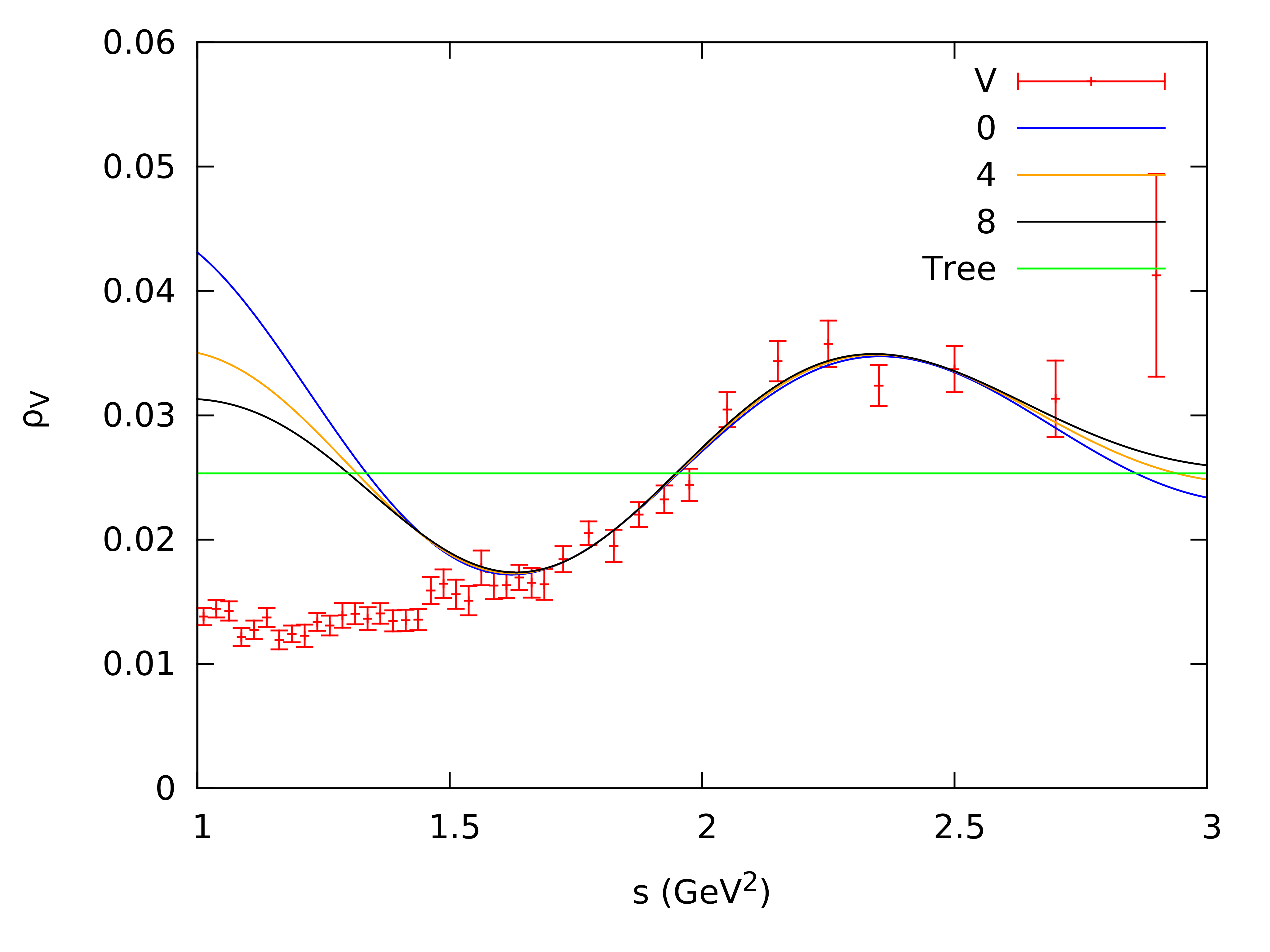}}
\vspace*{8pt}
\caption{Vector spectral function $\rho_V^{\protect\phantom{()}}(s)$, fitted
with the ansatz \eqn{eq:DVparam}, for different values of $\lambda_V= 0, 4, 8$, compared with the data points.\protect\cite{Pich:2016bdg}}
\label{fig:spectral-n}
\end{figure}

As soon as one moves from the region where the spectral function has been fitted, the model strongly deviates from the data. This is shown in Fig.~\ref{fig:spectral-n} which compares the measured spectral function with the fitted ansatz, for three different values of $\lambda_V=0,4,8$.
%
%
All models reproduce well $\rho_V^{\protect\phantom{()}}(s)$ in the fitted region ($s\ge 1.55\;\mathrm{GeV}^2$), but they fail badly below it. The default model ($\lambda_V=0$) assumed in Ref.~\refcite{Boito:2014sta} is clearly the worse one. When the power $\lambda_V$ is increased, the ansatz is able to slightly approach the data below the fitted range, while the exponential parameters $\delta_V$ and $\gamma_V$ adapt themselves to compensate the growing at high values of $s$ with the net result of a smaller duality-violation correction. The statistical quality of the fit improves also with growing values of $\lambda_V$, as shown in Table~\ref{tab:models} that gives the fitted parameters for different models ($\lambda_V=0,\cdots,8$), taking always the reference point $\hat s_0= 1.55\;\mathrm{GeV}^2$.

\begin{table}[t]
\tbl{Fitted values of $\alpha_s(m_\tau^2)$, in FOPT, and the spectral function ansatz parameters in Eq.~\eqn{eq:DVparam} with $\hat s_0= 1.55\;\mathrm{GeV}^2$, for different values of the power $\lambda_V$.\protect\cite{Pich:2016bdg}}
{\begin{tabular}{@{}ccccccc@{}}\toprule
$\lambda_V$  & $\alpha_{s}(m_{\tau}^{2})$ & $\delta_V$      & $\gamma_V$      & $\alpha_V$       & $\beta_V$
& p-value (\% )
\\ \colrule
0  & $0.298 \pm 0.010$          & $3.6 \pm 0.5$ & $0.6 \pm 0.3$ & $-2.3 \pm 0.9$ & $4.3 \pm 0.5$
& 5.3
\\ \hline
1  & $0.300 \pm 0.012$          & $3.3 \pm 0.5$ & $1.1 \pm 0.3$ & $-2.2 \pm 1.0$ & $4.2 \pm 0.5$
& 5.7
\\ \hline
2  & $0.302 \pm 0.011$          & $2.9 \pm 0.5$ & $1.6 \pm 0.3$ & $-2.2 \pm 0.9$ & $4.2 \pm 0.5$
& 6.0
\\ \hline
4  & $0.306 \pm 0.013$          & $2.3 \pm 0.5$ & $2.6 \pm 0.3$ & $-1.9 \pm 0.9$ & $4.1 \pm 0.5$
& 6.6
\\ \hline
8  & $0.314 \pm 0.015$          & $1.0 \pm 0.5$ & $4.6 \pm 0.3$ & $-1.5 \pm 1.1$ & $3.9 \pm 0.6$
& 7.7
\\ \botrule
\end{tabular}}
\label{tab:models}
\end{table}

From Table~\ref{tab:models}, one immediately realizes that there is a strong correlation between $\alpha_s(m_\tau^2)$ and the assumed model. This should not be a surprise, because we are just fitting models to data without any strong theoretical guidance (the OPE is no longer valid), and $\alpha_s$ has been converted into one more model parameter. Nevertheless, in spite of all caveats, one gets still quite reasonable values of the strong coupling. The actual uncertainties are much larger than the very naive fit errors shown in the table, since they totally ignore the strong instabilities appearing as soon as one moves from the selected point $\hat s_0= 1.55\;\mathrm{GeV}^2$. For instance, for the default $\lambda_V=0$ model, the fluctuations of $\alpha_s(m_\tau^2)$ in the interval $\hat s_0\in [1.15 , 1.75]~\mathrm{GeV}^2$ would force to increase by a factor of three the
error quoted in Table~\ref{tab:models}.\cite{Pich:2016bdg}
As the fit quality improves with growing values of $\lambda_V$, the fitted central values of $\alpha_s(m_\tau^2)$ approach also the much more solid determinations quoted in Table~\ref{tab:summary}.

The conclusion of this interesting exercise is obvious. The fitted values of $\alpha_s(m_\tau^2)$ extracted with this method strongly depend on the assumed spectral function model. Therefore, they are unreliable.
The slightly lower value of the strong coupling claimed in  Ref.~\refcite{Boito:2014sta} is just a consequence of their particular choice of model and has little to do with QCD; moreover, even if one believed its functional form, the uncertainties are grossly underestimated.

\section{Summary}

The results quoted in Table~\ref{tab:summary} are very robust, exhibiting a very good stability under small variations of the fit procedures, and they are rooted in solid theoretical principles
(except perhaps the one from the $s_0$ dependence, which assumes local duality). The overall agreement among determinations extracted under very different assumptions clearly shows their reliability and even indicates that our uncertainties are probably too conservative.
Averaging the five determinations, but keeping the smaller uncertainties to account for the large correlations, one finds
\be
\ba{c}
\alpha_{s}(m_\tau^2)^{\mathrm{CIPT}} \; =\; 0.335 \pm 0.013\, ,
\\[5pt]
\alpha_{s}(m_\tau^2)^{\mathrm{FOPT}} \; =\; 0.320 \pm 0.012\, .
\ea
\label{FinalValues}
\ee
The same results are obtained irrespective or whether one includes or not in the average the determination from the $s_0$ dependence of the moments. Averaging the CIPT and FOPT ``averages'' in Table~\ref{tab:summary}, we quote as our final determination of the strong coupling
\be
\alpha_{s}(m_\tau^2) \; =\; 0.328 \pm 0.012 \, .
\ee
These results nicely agree with the value of the strong coupling extracted\cite{Pich:2013lsa} from $R_\tau$.

After evolution up to the scale $M_Z$, the strong coupling decreases to
\be
\alpha_{s}^{(n_f=5)}(M_Z^{2})\; =\; 0.1197\pm 0.0014 \, ,
\ee
in excellent agreement with the direct measurement at the $Z$ peak from the $Z$ hadronic width,\cite{Agashe:2014kda}
$\alpha_{s}(M_Z^{2})\; =\; 0.1197\pm 0.0028$.
The comparison of these two determinations provides a beautiful test of the predicted QCD running; {\it i.e.} a very significant experimental verification of asymptotic freedom:
\be
\left.\alpha_{s}^{(n_f=5)}(M_Z^{2})\right|_\tau - \left.\alpha_{s}^{(n_f=5)}(M_Z^{2})\right|_Z
\; =\; 0.0000\pm 0.0014_\tau\pm 0.0028_Z\, .
\ee

Improvements on the determination of $\alpha_{s}(m_\tau^2)$ from $\tau$ decay data would require high-precision measurements of the spectral functions, specially in the higher kinematically-allowed energy bins. Both higher statistics and a good control of experimental systematics are needed, which could be possible at the forthcoming Belle-II experiment. On the theoretical side, one needs an improved understanding of higher-order perturbative corrections.

\section*{Acknowledgments}

A. Pich would like to express his gratitude to the Mainz Institute for Theoretical Physics (MITP) for its hospitality and the generous support to this workshop.
We also thank Michel Davier, Andreas Hoecker, Bogdan Malaescu, Changzheng Yuan and
Zhiqing Zhang for making publicly available the updated ALEPH spectral functions, with all
the necessary details about error correlations.
This work has been supported in part by the Spanish Government and ERDF funds from
the EU Commission [Grants No. FPA2014-53631-C2-1-P and FPU14/02990], by the Spanish
Centro de Excelencia Severo Ochoa Programme [Grant SEV-2014-0398] and by the Generalitat Valenciana [PrometeoII/2013/007].


\begin{thebibliography}{0}

\bibitem{Pich:2016bdg}
  A.~Pich and A.~Rodr\'{\i}guez-S\'anchez,
  arXiv:1605.06830 [hep-ph].


\bibitem{Pich:2013lsa}
A. Pich, 
{\it Prog. Part. Nucl. Phys.} {\bf 75}, 41 (2014).

\bibitem{Braaten:1991qm}
E.~Braaten, S. Narison and A.~Pich,
{\it Nucl. Phys. B} {\bf 373}, 581 (1992).

\bibitem{Pich:2015ivv}
A. Pich, 
in {\it High-precision $\alpha_s$ measurements from LHC to FCC-ee},
 pp. 37 (2015).

\bibitem{Pich:2013sqa}
A. Pich,
PoS ConfinementX {\bf 022} (2012).

\bibitem{d'Enterria:2015toz}
D. d'Enterria and P.~Z. Skands, editors,
{\it High-Precision $\alpha_s$ Measurements from LHC to FCC-ee},
arXiv:1512.05194.

\bibitem{Agashe:2014kda}
K.~A. Olive {\it et~al.},
{\it Review of Particle Physics},
{\it Chin. Phys. C} {\bf 38}, 090001 (2014).

\bibitem{Deur:2016tte}
  A.~Deur, S.~J.~Brodsky and G.~F.~de Teramond,
  arXiv:1604.08082 [hep-ph].

\bibitem{LeDiberder:1992zhd}
F.~Le~Diberder and A.~Pich,
{\it Phys. Lett. B} {\bf 289}, 165 (1992).

\bibitem{Davier:2013sfa}
M. Davier {\it et al.},  
{\it Eur. Phys. J. C} {\bf 74}, 2803 (2014).




\bibitem{Narison:1988ni}
S. Narison and A.~Pich,
{\it Phys. Lett. B} {\bf 211}, 183 (1988).

\bibitem{Braaten:1988hc}
E.~Braaten,
{\it Phys. Rev. Lett.} {\bf 60}, 1606 (1988);
%
{\it Phys. Rev. D} {\bf 39}, 1458 (1989).


\bibitem{Marciano:1988vm}
W.~J. Marciano and A.~Sirlin,
{\it Phys. Rev. Lett.} {\bf 61}, 1815 (1988).

\bibitem{Braaten:1990ef}
E. Braaten and C.-S. Li,
{\it Phys. Rev. D} {\bf 42}, 3888 (1990).

\bibitem{Erler:2002mv}
J. Erler,
{\it Rev. Mex. Fis.} {\bf 50}, 200 (2004).



\bibitem{Baikov:2008jh}
P.~A. Baikov, K.~G. Chetyrkin and J.~H. Kuhn,
{\it Phys. Rev. Lett.} {\bf 101}, 012002 (2008).

\bibitem{LeDiberder:1992jjr}
F.~Le~Diberder and A.~Pich,
{\it Phys. Lett. B} {\bf 286}, 147 (1992).

\bibitem{Pich:1999hc}
A. Pich and J. Prades,
{\it JHEP} {\bf 10}, 004 (1999);
%
{\bf 06}, 013 (1998).

\bibitem{Pivovarov:1991rh}
A.~A. Pivovarov,
{\it Z. Phys. C} {\bf 53}, 461 (1992). 

\bibitem{Davier:2008sk}
M.~Davier {\it et al.}, 
{\it Eur. Phys. J. C} {\bf 56}, 305 (2008);
%
{\it Rev. Mod. Phys.} {\bf 78}, 1043 (2006).

\bibitem{Schael:2005am} ALEPH Collaboration,
{\it Phys. Rept.} {\bf 421}, 191 (2005);
%
{\it Eur. Phys. J. C} {\bf 4}, 409 (1998);
%
{\it Phys. Lett. B} {\bf 307}, 209 (1993).

\bibitem{Ackerstaff:1998yj} OPAL Collaboration,
{\it Eur. Phys. J. C} {\bf 7}, 571 (1999).

\bibitem{Coan:1995nk} CLEO  Collaboration,
{\it Phys. Lett. B} {\bf 356}, 580 (1995).

%
\bibitem{Pich:2011bb}
A. Pich, arXiv:1107.1123.


\bibitem{Boito:2014sta}
D. Boito {\it et al.}, 
{\it Phys. Rev. D} {\bf 91}, 034003 (2015);
%
{\bf 85}, 093015 (2012);
%
{\bf 84}, 113006 (2011).


\bibitem{Cata:2008ye}
O. Cata, M. Golterman and S. Peris,
{\it Phys. Rev. D} {\bf 77}, 093006 (2008);
%
{\bf 79}, 053002 (2009);
%
{\it JHEP} {\bf 08}, 076 (2005).


\bibitem{Chibisov:1996wf}
B. Chibisov {\it et al.},  
{\it Int. J. Mod. Phys. A} {\bf 12}, 2075 (1997).


\bibitem{GonzalezAlonso:2010rn}
M. Gonz\'alez-Alonso, A. Pich and J. Prades,
{\it Phys. Rev. D} {\bf 81}, 074007 (2010);
%
{\bf 82}, 014019 (2010).

\bibitem{Rodriguez-Sanchez:2016jvw}
A.~Rodr\'{\i}guez-S\'anchez, M.~Gonz\'alez-Alonso and A.~Pich,
 arXiv:1602.06112.





\end{thebibliography}
\end{document}